\documentclass[preprint,nopacs,amsmath,amssymb]{revtex4-2}

\usepackage[dvips]{graphicx}
\usepackage{grffile}
\usepackage[normalem]{ulem}
\usepackage{dcolumn}
\usepackage{bm}
\usepackage{empheq}
\usepackage{color}

\newcommand{\av}[1]{\langle{#1}\rangle}

\newcommand{\bi}[1]{\mbox{\boldmath$#1$}}
\newcommand{\pp}[2]{\frac{\partial {#1}}{\partial {#2}}}

\begin{document}
\title{Emergent elasticity linked to topological phase transitions controlled via molecular chirality and steric anisotropy}
\author{Kyohei Takae}
\email{takae@iis.u-tokyo.ac.jp}
\affiliation{Department of Fundamental Engineering, Institute of Industrial Science, University of Tokyo, 4-6-1 Komaba, Meguro-ku, Tokyo 153-8505, Japan}
\author{Takeshi Kawasaki}
\email{kawasaki@r.phys.nagoya-u.ac.jp}
\affiliation{Department of Physics, Nagoya University, Nagoya 464-8602, Japan}
\date{\today}

\begin{abstract}
{\bf
Self-organisation into spatially modulated structures
has different nature from phase transition into uniform states. 
Skyrmions and half-skyrmions (merons) are representatives of such structures and are utilised in designing magnetoelectric, optical, and mechanoresponsive materials by controlling topological phases. 
However, skyrmions and half-skyrmions in molecular solids are rarely studied, though there is a universality in theoretical descriptions between magnetic and molecular systems with chiral interactions. 
Here we develop a simple physical system for controlling topological phases in a solid with chirality. 
We reveal that emergence of elastic fields from anisotropic steric interactions and intermolecular twisting is a key to control helical and half-skyrmion structures. 
Utilising the coupling between the emergent elastic fields and molecular orientations, we successfully control topological phases by temperature, external electromagnetic fields, and anisotropic stresses. 
The concept of the emergent elasticity provides a control system for designing molecular and macromolecular solids with tunable electro- and magneto-mechanical properties. 
}
\end{abstract}

\maketitle

\clearpage


Understanding and controlling phase transitions into spatially modulated structures such as helices are of importance both in fundamental physics and technological applications. It is because they exhibit different nature from that of phase transitions into uniform states which has been established~\cite{Brazovskii}, and materials properties can be controlled mesoscopically by utilising non-uniform microphases~\cite{Leibler}. Skyrmions and half-skyrmions (merons) are representative objects of such structures and are of interest in magnetic systems~\cite{NagaosaTokura2013,Kawamura}, Bose-Einstein condensates~\cite{Ho-BEC}, quantum Hall systems~\cite{QHE}, dielectrics~\cite{Nahas-polar,Ramesh-polar2019}, liquid crystals~\cite{Fukuda2011,Fukuda2017}, and active matter~\cite{DePablo,Shankar-active}. These objects are manipulated in designing magnetoelectric, optical, and mechanoresponsive materials~\cite{Tokura2021review,Smalyukh,pinning}. In liquid crystals, in particular, molecular chirality plays an important role on the formation of helical and half-skyrmion phases~\cite{deGennesProst,Mermin,Tschierske2014,Smalyukh}. 
In theoretical description of chiral liquid crystals, a spontaneous twisting term reflecting chirality is incorporated to examine formation of mesoscale cooperative structures~\cite{deGennesProst,Mermin,Fukuda2011,Yeomans}. This term resembles Dzyaloshinskii-Moriya interaction in chiral magnets~\cite{Roessler,NagaosaTokura2013}, incorporating liquid crystalline symmetry of the order parameter. Thus the formation of helical and half-skyrmion phases in cholesteric liquid crystals is explained~\cite{Fukuda2011}.
Recently, furthermore, half-skyrmion phases in metal-organic frameworks (MOFs) were theoretically proposed using a similar model~\cite{Skyrmion-MOF}. Therefore, it is reasonable to expect the existence of similar phase transitions in other chiral molecular systems such as organic crystals~\cite{Viedma2006,softcrystal}, colloidal crystals~\cite{Dogic2017,Dijkstra2016}, and biological systems~\cite{chemistry} by utilising the same theoretical framework,
though the existence of half-skyrmion structures in these substances has not yet been investigated. 

However, there is a crucial difference between liquid crystals and crystals. In molecular crystals and in MOFs, change of molecular configurations associated with phase transitions induces lattice distortion, resulting in the emergence of elastic field, though the role of the elastic field remains elusive. We call this elastic field an {\it emergent elastic field}, which is absent in cholesteric liquid crystals due to its liquid nature in molecular translation. This is an analogue of the emergent electromagnetic field in condensed matter, which is known to play an essential role on manipulating skyrmions~\cite{emergent,Nagaosa-emergent}. Therefore, it is of crucial importance to understand how the emergent elastic field is linked to topological phase transitions in molecular crystals.

Here, we reveal an essential role of the emergent elastic field associated with topological phase transitions into helical and half-skyrmion phases in a solid. To simplify the problem, we develop a molecular dynamics model incorporating the coupling between molecular orientation and crystal elasticity by assuming that molecules have spheroidal shape with chirality. The spheroidal molecule has anisotropic steric interaction, hence molecular rotation induces lattice distortion. By controlling the steric anisotropy and twisting interaction between adjacent molecules, phase transitions into helical and half-skyrmion phases in a solid are realised. Long-range nature of elastic correlation is important to control phase behaviour and domain formations. As a consequence of this elastic coupling, these phases can be switched by varying temperature and by applying external electromagnetic fields and anisotropic stresses. Our work reveals a link between topological phase transitions and the elastic fields, which provides a control system for designing molecular and macromolecular solids with tunable electro- and magneto-mechanical properties.

\begin{figure*}[t!]
\centering
\includegraphics[width=15cm]{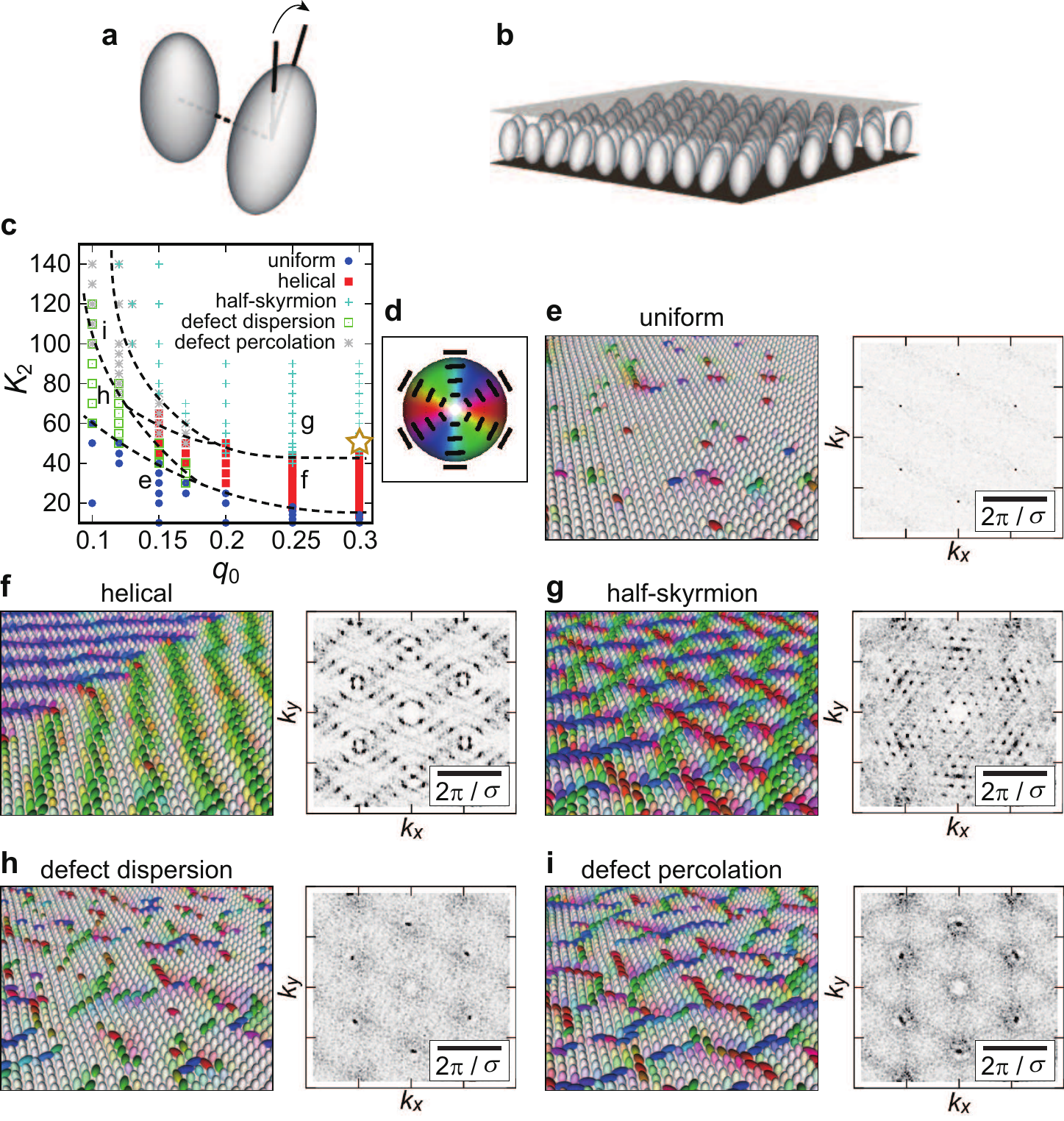}
\caption{
{\bf Phase behaviour in a monolayer geometry.}
{\bf a.} Molecular twisting in our model. The favoured twist angle is given by $q_0$, and the rigidity of the twist is determined by $K_2$ (see Methods for this definition).
{\bf b.} Monolayer geometry of this study (see Methods for detail).
{\bf c.} Phase diagram with respect to $q_0$ and $K_2$ at low temperatures. The phase boundaries are shown for reference. The star symbol denotes the phase point examined in Fig.\ref{fig:half-skyrmion}.
{\bf d.} Colour notation for the snapshots in {\bf e-i} and the following figures.
{\bf e-i.} Diagonal view of the real spatial structure and corresponding structure factor (see Methods for this definition) at each phase.
The temperature $T=0.05$ in this figure.
}
\label{fig:phase}
\end{figure*}

\vspace{5mm}
\noindent{\bf Phase diagram with respect to twist parameters}\\
We construct a molecular dynamics model by incorporating twisting interactions between adjacent molecules, which is schematically described in Fig.\ref{fig:phase}a.  The alignment of neighbouring molecules is stable when they exhibit a certain twist angle determined by $q_0$, and the rigidity of this twist is given by $K_2$ (see Methods for this definitions of these parameters). These molecules are confined to realise a monolayer geometry, as shown in Fig.\ref{fig:phase}b, forming two-dimensional crystals (see Extended Data Figure \ref{fig:Q2d} for larger thickness geometries). Using this model system, we demonstrate phase controllability via the manipulation of $q_0$ and $K_2$ at low temperatures, as displayed in Fig.\ref{fig:phase}c.
For small $K_2$ and $q_0$ values, almost all the molecules are oriented normal to the layer to form a uniform phase, as displayed in Fig.\ref{fig:phase}e. In this state, the structure factor (see Methods for this definition) exhibits only the Bragg peaks representing the nearest neighbour particles, and no mesoscopic structure forms. As the twist rigidity $K_2$ is increased, the number of molecules that orient tangential to the monolayer increases.
For large $q_0$ values, long-range orders of helical and half-skyrmion states are formed, as displayed in Fig.\ref{fig:phase}f and g, respectively.
Conversely, for small $q_0$ values ($\sim 0.1$), particle orientation does not vary continuously in space and orientational defects are dispersed heterogeneously, exhibiting a halo structure at low wavenumbers, as shown in Fig.\ref{fig:phase}h. As $K_2$ is increased, the number of defects increases, and eventually results in the division of ordered particles into compartments (Fig.\ref{fig:phase}i).
The transitions between these structures are gradual except for the sharp transition between the helical and the half-skyrmion phase (see Extended Data Figure \ref{fig:q03}).
Therefore, we successfully control the topological structure of this system by controlling the twisting interactions.

\begin{figure}[t!]
\centering
\includegraphics[width=15cm]{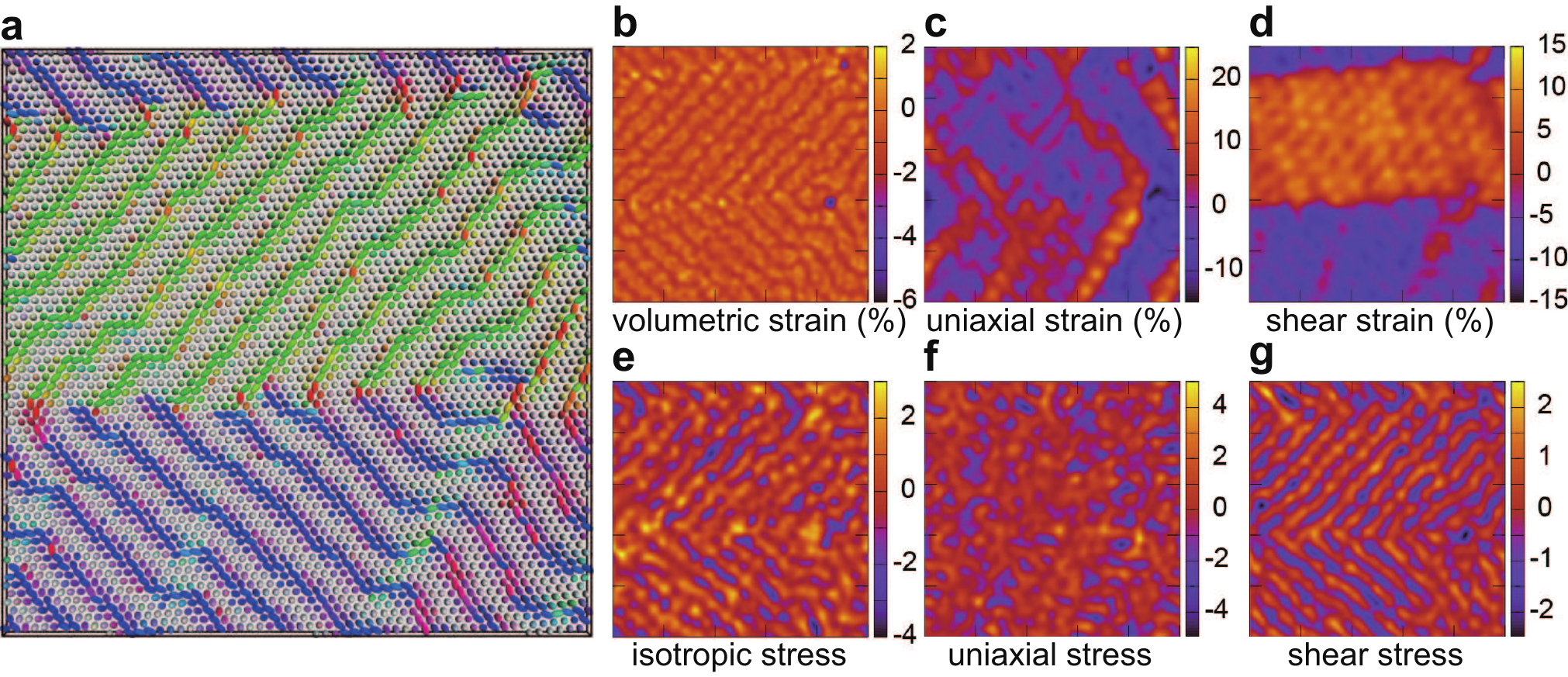}
\caption{
{\bf Emergent elastic field of a helical state.}
{\bf a.} Particle configuration corresponding to Fig.\ref{fig:phase}f, wherein the system is divided into two major domains with different particle orientations (indicated in blue and green in this figure).
{\bf b-d.} Strain field of this configuration (see Methods for this definition). The volumetric strain represents the volume dilation and compression and is small. The emergence of a large macroscopic uniaxial strain in {\bf c} and a large shear strain in {\bf d} result from the formation of a helical state, wherein tangentially aligned particles produce large strains along their orientations.
{\bf e-g.} Stress field of this configuration (see Methods for this definition). ({\bf e}) Isotropic stress corresponding to the volumetric strain via the bulk modulus (volume compressibility). ({\bf f}) Uniaxial stress and ({\bf g}) shear stress, which exhibit the same spatial pattern as the particle configuration, unlike the strain correspondence.
}
\label{fig:Em-heli}
\end{figure}

\vspace{5mm}
\noindent{\bf Emergence of elastic fields in topological phases}\\
The characteristic feature of our system is the emergence of strain and stress fields. In this study, a two-dimensional solid state is realised at a high temperature because the average density of this system is large, and orientational phase ordering proceeds without particle migration. Because each molecule exhibits steric anisotropy, the tangential orientation of a molecule generates strain and stress around the molecule, which is similar to the generation of an elastic field with a long-range spatial correlation around an inclusion and a defect~\cite{Eshelby,Lemaitre}. As referred in the introduction, we call this elastic field an {\it emergent elastic field}.
The concept of the emergent elastic field can explain the domain formation shown in Fig.\ref{fig:phase}f, wherein the helical state is divided into domains with different wave vectors, whose angles are approximately $\pm 30^\circ$. In Fig.\ref{fig:Em-heli}, we present the emergent strain and stress fields (see Methods for this definition). As shown in Fig.\ref{fig:Em-heli}c and d, a large anisotropic strain is induced inhomogeneously because molecular orientations in different helical domains produce different strain components. If a uniform helical structure is formed, it produces large uniform strain which generate large anisotropic pressure, thus becomes mechanically unstable. By forming domain structures, the anisotropic stress is macroscopically reduced and localised as shown in Fig.\ref{fig:Em-heli}f and g. Such domain formation does not occur for the half-skyrmion state shown in Fig.\ref{fig:phase}g because half-skyrmions form an isotropic hexagonal solid at low temperatures (see Extended Data Figure \ref{fig:Em-uni} and \ref{fig:Em-sky} for the uniform and half-skyrmion phases, respectively).
The emergent elastic field is not relevant in cholesteric liquid crystals due to its liquid nature in molecular translation, but it can be important in chiral smectics~\cite{deGennesProst} and elastomers~\cite{Terentjev} because molecular rotations induce layer compression and network deformation, respectively. The investigation of (half-)skyrmion structures in these substances and in molecular solids should be conducted to reveal the impact of the emergent elastic fields.


\begin{figure}[t!]
\centering
\includegraphics[width=14cm]{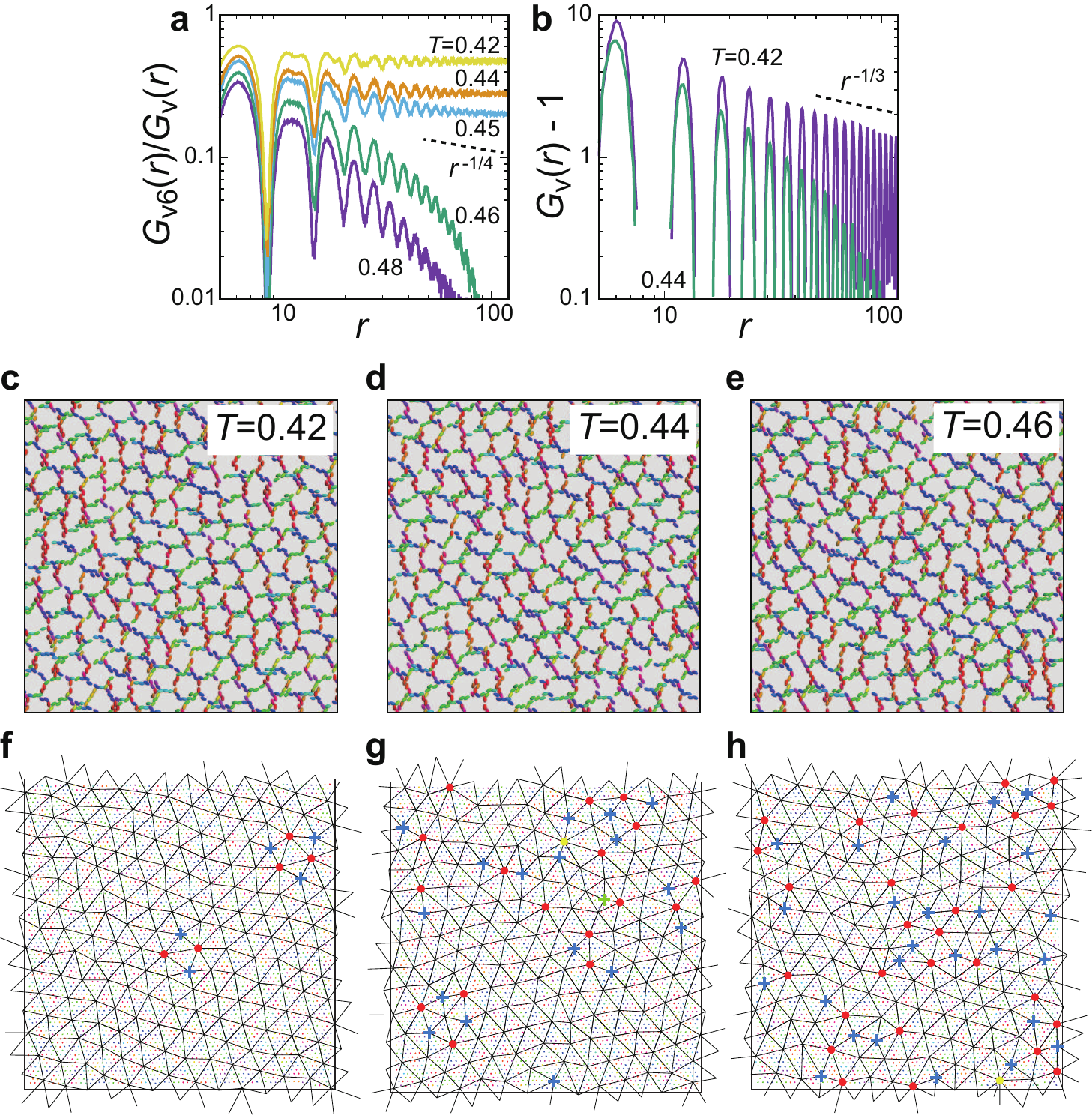}
\caption{
{\bf Thermal stability of the half-skyrmion phase.}
{\bf a.} Hexatic correlation function divided by the radial distribution function (see Methods for this definition) of the vortices. The hexatic correlation increases as the temperature decreases via the characteristic power-law decay.
{\bf b.} Spatial correlation function of the half-skyrmion cores. It exhibits a power-law decay in the solid phase, whereas it decays exponentially in the hexatic and liquid phases.
{\bf c-e.} Real spatial structures at $T=0.42$ ({\bf c}: half-skyrmion crystal), $T=0.44$ ({\bf d}: half-skyrmion hexatic), and $T=0.46$ ({\bf e}: half-skyrmion liquid). We only display particles with $n_z<1/\sqrt{2}$.
{\bf f-h.} Corresponding Delaunay triangulation. The red (yellow) circles denote 7(8)-member coordinated vortices, whereas the blue (green) crosses denote 5(4)-member coordinated vortices.
$K_2=50$ and $q_0=0.3$ in this figure, which are indicated by the star symbol in Fig.\ref{fig:phase}c.
}
\label{fig:half-skyrmion}
\end{figure}

\vspace{5mm}
\noindent{\bf Thermal phase transitions of the helical and the half-skyrmion phases}\\
The phase diagram shown in Fig.\ref{fig:phase} indicates the stable structure at a low temperature. In previous studies, the phase transitions between topological phases were often examined by varying the temperature and external perturbations such as the magnetic field. In this section, we investigate the thermal stability of the helical and half-skyrmion phases. For the former phase, we demonstrate the thermal hysteresis of the helical phase in Extended Data Figure \ref{fig:helical}. As the temperature is increased, the helical state transforms into a half-skyrmion state, which is indicated by the abrupt change in the potential energy of the system. This phase transition is reversible and exhibits a hysteresis loop, implying its first-order nature.
For the latter phase, the vortex structure in half-skyrmion phase is stable over wide temperature range. However, another phase transition with respect to the long-range ordering of half-skyrmions emerges, where the transitions between the liquid, hexatic, and crystalline half-skyrmions occur as shown in Fig.\ref{fig:half-skyrmion}. The positional and bond-orientational ordering of the half-skyrmions exhibits a characteristic two-dimensional melting behaviour based on Kosterlitz-Thouless-Halperin-Nelson-Young (KTHNY) scenario ~\cite{KT1973,NelsonHalperin,Young,Krauth2015,Huang,Zazvorka}. This is confirmed by the correlation functions of the half-skyrmions, as displayed in Fig.\ref{fig:half-skyrmion}a and b. The hexatic correlation function shown in Fig.\ref{fig:half-skyrmion}a (see Methods for this definition) decays exponentially for $T \ge 0.46$ (characteristic of the liquid phase), algebraically for $T\simeq 0.45$ with an exponent close to $-1/4$ (characteristic of the hexatic phase), and approaches a constant value for $T \le 0.42$ (characteristic of the solid phase).
The hexatic and solid phases is distinguished by the decay of the pair correlation function along with the direction of the bond orientation angle (see Methods for further details), which is displayed in Fig.\ref{fig:half-skyrmion}b. The pair correlation function decays exponentially for $T=0.44$ (hexatic phase), whereas it exhibits a power-law decay at $T=0.42$ with an exponent near $-1/3$, which is characteristic of the stability limit of a solid phase in the two-dimensional melting theory.
The real-space topological structures are presented in Fig.\ref{fig:half-skyrmion}c-h, wherein the filtered particle configurations and corresponding Delaunay triangulation results are displayed in upper (c-e) and lower (f-h) rows, respectively. In the two-dimensional melting theory~\cite{KT1973,NelsonHalperin,Young},
only dislocation pairs with a zero Burgers vector are excited in the solid phase. They separate into single dislocations (disclination pairs) in the hexatic phase, and the liquid phase is characterised by the disclination-unbound phase in which isolated point defects disperse. This characteristic feature is captured in the Delaunay triangulation. Our results based on KTHNY scenario are consistent with those found recently in magnetic skyrmion systems~\cite{Huang,Zazvorka}.


\begin{figure*}[t!]
\centering
\includegraphics[width=11.9cm]{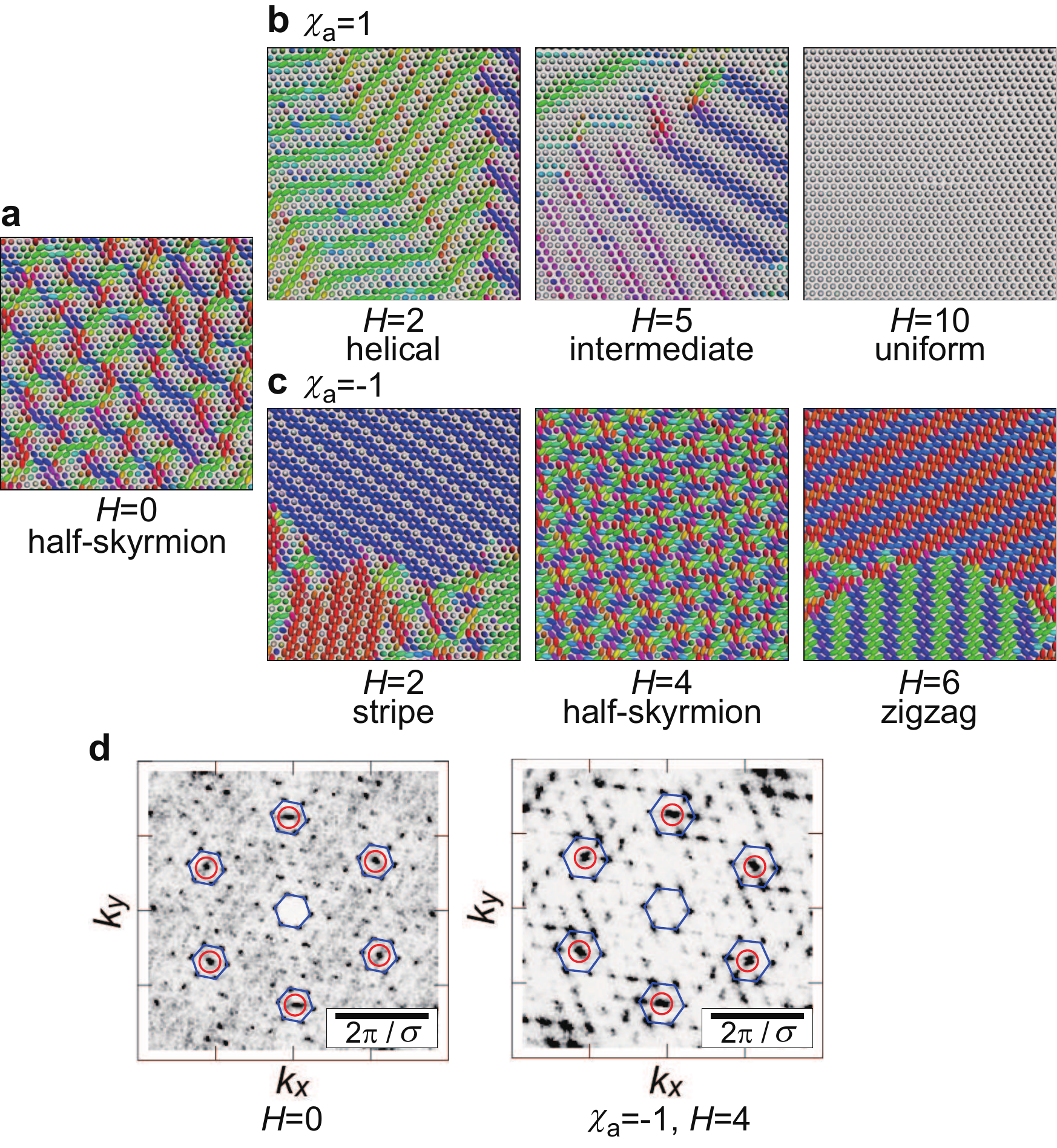}
\caption{
{\bf Response to external fields under the field-cooling condition.}
{\bf a-c.} The external field effects on the half-skyrmion phase ({\bf a}) with positive ({\bf b}) and negative ({\bf c}) magnetic anisotropy are shown, where the magnetic anisotropy is defined by $\chi_{\rm a}=\chi_\parallel-\chi_\perp$, and $H$ denotes the magnitude of the external magnetic field applied normal to the monolayer.
{\bf d.} The structure factors obtained at $H=0$ and $H=4$ with negative anisotropy implies that the topology of these structures is the same. The red circles and blue hexagons represent the Bragg spots corresponding to the nearest molecules and half-skyrmions, respectively, indicating that the distance between the half-skyrmion cores becomes smaller.
$K_2=50$, $q_0=0.3$, and $T=0.05$ in this figure.
}
\label{fig:field}
\end{figure*}

\vspace{5mm}
\noindent{\bf Phase switching and re-entrant phase transition by an external field}\\
Another method for controlling topological structures is to apply an external magnetic/electric field to substances, as is often examined in magnetic skyrmion systems~\cite{NagaosaTokura2013} and liquid crystals~\cite{deGennesProst}. In this paper we examine magnetic field effects for our convenience, but the same conclusion holds for electric field effects, because static field effects without impurities are considered. In rod-like uniaxial molecular systems, the magnetic susceptibility $\tensor{\chi}$ exhibits both $\chi_\parallel > \chi_\perp$ (positive anisotropy) and $\chi_\parallel < \chi_\perp$ (negative anisotropy), where $\chi_\parallel$ ($\chi_\perp$) is the susceptibility of a molecule parallel (perpendicular) to the molecular long axis, respectively~\cite{deGennesProst}. For positive (negative) anisotropy, the orientation of molecular long axis becomes parallel (perpendicular) to the external field.
This feature is reminiscent of the perpendicular magnetic anisotropy systems in transition-metal oxides~\cite{PMA}, in which the magnetic easy axis is tangential to the thin films.
In this study, we neglect the dipole-dipole interaction between the molecules. Then the molecular response to the external field becomes second-order, which is often assumed in liquid crystals (see Methods for further detail)~\cite{deGennesProst}. 
We present the field response of a half-skyrmion phase in Fig.\ref{fig:field}, where $H$ denotes the magnitude of the field applied normal to the monolayer (see Extended Data Figure \ref{fig:10} and \ref{fig:20} for other phases).
Starting from the half-skyrmion state shown in Fig.\ref{fig:field}a, molecules with positive (negative) anisotropy begin to align parallel (perpendicular) to the external field. For the positive anisotropy case shown in Fig.\ref{fig:field}b, molecules at the edges of the half-skyrmions change their orientation to align parallel to the external field, resulting in the coalescence of half-skyrmions to form helical structures ($H=2$). By further increasing this field, an increasing amount of particles align parallel to the external field, and uniformly aligned state is eventually realised at $H=10$. The transition from the helical state to the uniform state is similar to the field responses of cholesteric liquid crystals, wherein uniformly aligned domains are separated by sharply twisting walls during the intermediate stage~\cite{deGennesProst} (see Fig.6.13 in their book). Therefore, the transition pathway for positive anisotropy demonstrated by our system is reasonable, while it sharply contrasts with magnetic skyrmion systems in which the helical state transforms into the skyrmion state under the application of an external magnetic field.

For negative anisotropy systems, conversely, a curious phase transformation emerges, as displayed in Fig.\ref{fig:field}c. The initial half-skyrmion state transforms into a helical-like striped structure with a small pitch under a weak external field ($H=2$). Subsequently, another half-skyrmion state appears at $H=4$, which finally transforms into a two-dimensional zig-zag structure at $H=6$ (the morphology of the two-dimensional structure depends on the twist rigidity $K_2$, as shown in Extended Data Figures \ref{fig:10}, \ref{fig:20}, and \ref{fig:50}).
Although the real space structures of the half-skyrmion states in Fig.\ref{fig:field}a and c look quite different, their topological structures exhibit similar symmetries. This is confirmed by the structure factor displayed in Fig.\ref{fig:field}d. Both states have six-fold symmetrical peaks, as highlighted by the blue hexagons, which represent the spatial correlation of the vortices. The difference between these peaks is represented by the size of the hexagons, shown that the vortex size is reduced under the external field. This re-entrant phase transition is a unique feature of negative anisotropy systems, which implies that interactions between the steric and magnetic/dielectric anisotropy results in rich phase behaviour in molecular solids.
As described in Fig.\ref{fig:Em-heli}, an elastic field emerges via phase transformations, which implies that our system exhibits electro- and magneto-striction as a cross-coupling effect, which demonstrates the great potential of this system for the design of functional materials.


\begin{figure*}[t!]
\centering
\includegraphics[width=15cm]{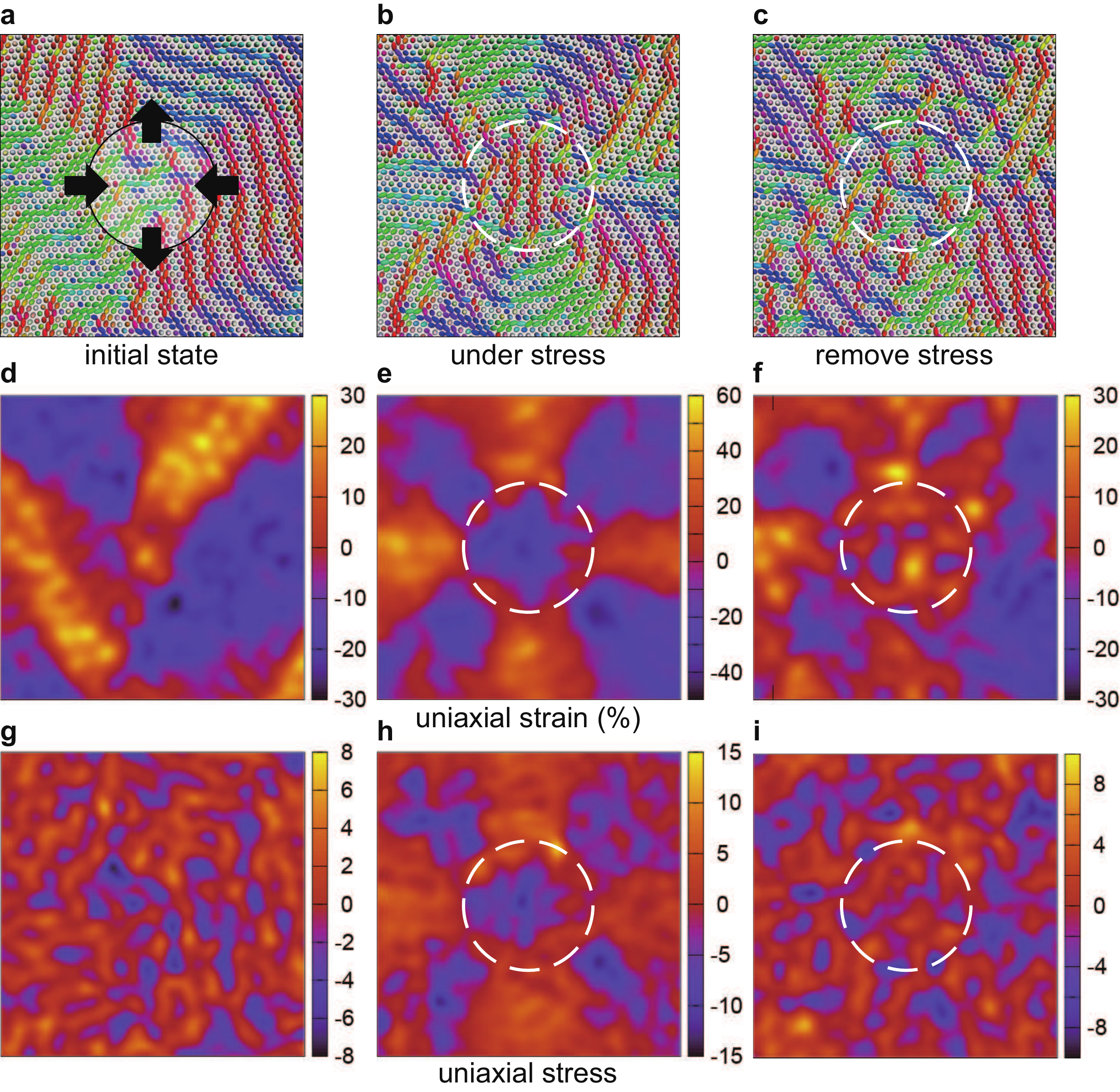}
\caption{
{\bf Response to local anisotropic stress.}
{\bf a.} The helical state for initial unstressed condition. The circle at the centre and the arrows represent the area and the orientation of applied anisotropic stress in ({\bf b}).
{\bf b.} Molecular configuration under the stress, where the molecular configuration is perturbed globally.
{\bf c.} The molecular configuration after stress removal. Half-skyrmion structure is maintained at the central area after the stress is relaxed.
{\bf d-f.} Uniaxial component of the emergent strain field for ({\bf d}) initial state, ({\bf e}) stressed state, and ({\bf f}) relaxed state.
{\bf g-i.} Uniaxial component of the emergent stress field for ({\bf g}) initial state, ({\bf h}) stressed state, and ({\bf i}) relaxed state.
$K_2=40$, $q_0=0.3$, and $T=0.2$ in this figure.
}
\label{fig:uniaxial}
\end{figure*}

\vspace{5mm}
\noindent{\bf Phase switching by local anisotropic stress}\\
Finally, we present a crucial role of the emergent elastic fields on phase controllability. As revealed in Fig.\ref{fig:Em-heli}, the helical domain and the emergent strain field exhibit the same spatial patterns. This suggests that domain orientation can be controlled by external anisotropic strain and stress. To see this, we display mechanical response of a helical state in Fig.\ref{fig:uniaxial}. When an external mechanical stress is applied locally (circled region in Fig.\ref{fig:uniaxial}a), both the helical pattern inside and outside the stressed region transform considerably, as displayed in Fig.\ref{fig:uniaxial}b (the white broken circle denotes the stressed region). For the former, the helical pitch becomes parallel to the compression direction, which indicates that molecules inside the stressed region orient to the elongation direction, reducing the uniaxial stress inside the circular region. For the latter, helices change their pitch direction parallel to dipole field. This is because both the strain and the stress field are induced outside the stressed region due to the long-range quadrupolar nature of elastic correlation, as displayed in Fig.\ref{fig:uniaxial}e and h. Both the strain field and the stress field exhibit angle dependent long-range correlation under the external stress, where they become positive along $0^\circ$ and $90^\circ$, and negative along $\pm 45^\circ$. After the external stress is removed, the helical state inside the circular region transforms into half-skyrmion structure whereas the helical domains outside the stressed region remain almost unchanged, as displayed in Fig.\ref{fig:uniaxial}c. The remnant strain in relaxed states also exhibits the same spatial heterogeneity as molecular orientation (Fig.\ref{fig:uniaxial}f), whereas the stress is localised (Fig.\ref{fig:uniaxial}i). Thus, we successfully control topological phases by applying mechanical stress.

\vspace{5mm}
\noindent{\bf Discussion and summary}\\
In this study, we succeeded to control the topological phase transition between the helical and the half-skyrmion phases. This transition is reversible without plastic deformation. This reversibility is attributed to the fact that the effective aspect ratio of our molecule is close to unity: lattice distortion induced by molecular rotation is not large so that crystal defects do not form.
When the particles with large aspect ratio such as fd-viruses~\cite{Dogic} and cellulose nanocrystals~\cite{MacLachlan} are utilised in a solid phase, particle rotation induces plastic deformation which results in irreversible phase transformation. Therefore, a particle should have a shape deformed from isotropic shape only slightly, and indeed it can be synthesised experimentally~\cite{Zerrouki,Singh,Nanorod,Rosi,Sacanna2021}. A densely packed solid thin-film of these particles will be candidates to exhibit phase transition between the helical and the half-skyrmion phase. Furthermore, it is possible to measure elastic heterogeneity by atomic force microscope for molecular and nanoparticle crystals~\cite{AFM}, and by confocal microscope for crystals composed of micrometre-scale colloidal particles~\cite{Spaepen}. These measurements can examine the relationship between the formation of topological phases and the emergent elastic field.

In summary, we presented a physical principle for topological phase control using material parameters and external electromagnetic fields in a model molecular solid. The most important aspect of our model is the emergent elastic field that is produced by the interactions between the molecular steric anisotropy and twisting interactions. We also presented a method for controlling the phases of this system using external fields, which is achieved due to the competition between the emergent elastic field and the paramagnetic/paraelectric response. For the first time, we identified a topologically re-entrant phase transition that is induced by an external field in negative anisotropy systems. In this paper, we considered the electromagnetic effects in a simple manner by neglecting the electromagnetic interactions between the induced dipoles. The inclusion of dipole-dipole interaction results in controllable polar orders with large mechanical responses~\cite{Takae2018antiferro}. This cross-coupling may also be a key to control emergent elastic fields associated with topological phase transitions induced by anisotropic mechanical stresses~\cite{Tokura2015strain,Wang2019-mechanical}.

\clearpage
\noindent{\bf \large Methods}

\noindent{\bf Molecular dynamics model}
We construct a simple molecular model exhibiting phase transitions into helical and half-skyrmion phases, by applying conventional knowledge in liquid crystals~\cite{deGennesProst}.
The potential energy of our system is given by
\begin{align}
U=&\sum_{i<j}4\epsilon(1+A_{ij}+B_{ij})\bigg(\frac{\sigma}{r_{ij}}\bigg)^{12} + U_{\rm ex} + U_{\rm wall}, \\
A_{ij}=&\eta[(\bi n_i\cdot \hat{\bi r}_{ij})^2+(\bi n_j\cdot \hat{\bi r}_{ij})^2],\\
B_{ij}=&\frac{K_2}{2}[(\bi n_i\cdot\bi n_j)(\bi n_i\times\bi n_j)\cdot\hat{\bi r}_{ij}-q_0]^2,
\end{align}
where $\epsilon$ and $\sigma$ denote the characteristic energy and length in our model, $r_{ij}=|\bi r_{ij}|$ and $\hat{\bi r}_{ij}=\bi r_{ij}/r_{ij}$ are the absolute value and unit vector of the intermolecular displacement, respectively, and $\bi n_i$ denotes the molecular orientation of the uniaxial molecules. $A_{ij}$ represents the symmetric steric repulsion used to mimic spheroidal molecules in a condensed phase~\cite{Takae2014origlass,Takae2017relaxor} in which the aspect ratio is $p=(1+2\eta)^{1/6}$ for small $\eta$ values. $B_{ij}$ represents twisting interactions arising from molecular chirality that adjacent molecules favour to align with a twist, where the favoured twist angle and twist rigidity are given by $q_0$ and $K_2$, respectively~\cite{Meer,Memmer}. 
This is the discretised form of the twist Frank energy $(K_2/2)(\bi n\cdot{\rm curl}\bi n +q_0)^2$ in liquid crystal theory~\cite{deGennesProst}, wherein the bilinear term is the same as the Dzyaloshinskii-Moriya interaction in magnetic systems ($D\bi n\cdot{\rm curl}\bi n$ with $D=K_2q_0$)~\cite{NagaosaTokura2013}.
In this paper, we assume that the intermolecular interaction has only the short-range repulsive steric term in order to examine the role of the twist interaction and the steric anisotropy in the simplest manner. We note that inclusion of van der Waals attractive interaction does not change the qualitative features of the formation of the half-skyrmion and helical phases.

$U_{\rm ex}$ represents the external field effects of the molecular orientation, which is defined as
\begin{equation}
U_{\rm ex}=-\sum_i \frac{\chi_{\rm a}}{2}(\bi n_i\cdot\bi H)^2,
\end{equation}
where $\bi H$ is the external field and $\chi_{\rm a}=\chi_\parallel-\chi_\perp$ denotes the anisotropic susceptibility. Here, $\chi_\parallel$ ($\chi_\perp$) is the susceptibility of a molecule that is parallel (perpendicular) to the molecular long axis. A molecule is oriented parallel (perpendicular) to the external field for positive (negative) $\chi_{\rm a}$ values. This form is often adopted for liquid crystals~\cite{deGennesProst}. A small number of molecules are known to exhibit negative magnetic anisotropy~\cite{Buka}, whereas various liquid crystalline molecules that exhibit negative dielectric anisotropy are found~\cite{deGennesProst}.
In numerical simulations, we normalise $\chi_{\rm a}=\pm 1$ by redefining $\bi H$ as $\sqrt{|\chi_{\rm a}|}\bi H$.

$U_{\rm wall}$ arises from the confinement due to the walls of the system. In this study, we assume a monolayer of spheroid-like molecules without surface anchoring to examine the impact of molecular twisting on the molecular configuration of the system in a simple manner. We therefore assume
\begin{equation}
U_{\rm wall}=\sum_i \epsilon[(\sigma/z_i)^{12}+(\sigma/(L_z-z_i))^{12}],
\end{equation}
where $z_i$ is the z-coordinate of $i$-th molecule and $L_z$ is the distance between the two walls. By utilising $L_z=2\sigma$, it can be easily confirmed that the molecules form a monolayer without undulation along the $z$-axis (see Extended data figure \ref{fig:Q2d} for larger $L_z$ values). $\hat{\bi r}_{ij} \perp z$ in this monolayer geometry, and hence $A_{ij}=0$ when two adjacent molecules align perpendicular to the intermolecular displacement $\bi n \perp \hat{\bi r}_{ij}$, while $B_{ij}=0$ when two molecules exhibit the twisting angle $\theta=(1/2)\sin^{-1}2q_0$.

The equations of motion with respect to the molecular position and orientation are given by
\begin{align}
m\ddot{\bi r}_i=&-\pp{U}{\bi r_i},\label{eq:trans}\\
I(\tensor{1}-\bi n_i\bi n_i)\cdot\ddot{\bi n}_i=&-(\tensor{1}-\bi n_i\bi n_i)\cdot\pp{U}{\bi n_i},
\label{eq:rot}
\end{align}
where $m$ is the molecular mass, $I$ is the moment of inertia with respect to the molecular long axis,
and $\tensor{1}$ is the unit tensor. By using $(1/2)d^2 |\bi n_i|^2/dt^2=|\dot{\bi n}_i|^2+\bi n_i\cdot\ddot{\bi n}_i=0$, Eq.(\ref{eq:rot}) may be rewritten as
\begin{equation}
I\ddot{\bi n}_i=-I\bi n_i|\dot{\bi n}_i|^2-(\tensor{1}-\bi n_i\bi n_i)\cdot\pp{U}{\bi n_i}.
\label{eq:rot2}
\end{equation}
Eq.(\ref{eq:trans}) and (\ref{eq:rot2}) are integrated during the time evolution under the NVT-ensemble using Nos\'e-Hoover thermostat~\cite{Allen}.

In this study, we assume densely packed systems. The volume fraction $\rho=\pi Np\sigma^3/6V$ is set to 0.3, where $N$ is the number of the particles and $V$ is the system volume.
Temperature $T$ is noted in the unit of $\epsilon/k_{\rm B}$ where $k_{\rm B}$ is the Boltzmann constant.
The stress tensor in Fig.\ref{fig:Em-uni} and \ref{fig:uniaxial} is noted in the unit of $\epsilon/\sigma^3$.
We adopt a periodic boundary condition in the $x$ and $y$ directions. For most numerical simulations, we choose $N=4000$, except for Fig.\ref{fig:half-skyrmion} ($N=64000$ to examine long-range correlations), Extended Data Figure \ref{fig:Q2d} ($N=16000$), and \ref{fig:helical} ($N=64000$).

\noindent{\bf Structure factor}
As displayed in the figures in the main text, the topology of a structure is characterised using the arrangement of the molecules that are aligned perpendicular to the $z$-axis. Therefore, we define a filtered density distribution $\rho_<(\bi r)=\sum_i\theta(n_{zi})\delta(\bi r-\bi r_i)$ and calculate the structure factor using $\rho_<(\bi r)$ to determine the topological structure, where $\theta(x)=1$ for $x<1/\sqrt{2}$ and $\theta(x)=0$ otherwise.

\noindent{\bf Emergent elastic fields}
In Fig.\ref{fig:Em-heli}, Extended Data Figure \ref{fig:Em-uni}, and \ref{fig:Em-sky}, we present local strain and stress fields. The strain tensor for each particle is defined by
\begin{equation}
\tensor{\varepsilon}_i=\frac{2}{r_{\rm M}^2N_{{\rm b}i}}\sum_j\bi r_{ij}\bi r_{ij},
\end{equation}
where $N_{{\rm b}i}$ is the coordination number of $i$-th particle in which the cutoff length is the first minimum of the radial distribution function. This sum is obtained over the coordinated particles, and $r_{\rm M}$ is the first maximum of the radial distribution function required to normalise the strain tensor such that $\tensor{\varepsilon}=\tensor{1}$ at the ground state. The volumetric, uniaxial, and shear strains are defined as $\det[ \tensor{\varepsilon}]-1$, $\varepsilon_{xx}-\varepsilon_{yy}$, and $\varepsilon_{xy}$, respectively. The strain field displayed in the figures is obtained by coarse-graining the strain tensor as
\begin{equation}
\tensor{\varepsilon}(\bi r)=\int d\bi r' w(\bi r-\bi r')\sum_i\tensor{\varepsilon}_i\delta(\bi r' -\bi r_i),
\end{equation}
where $w(\bi r)=(1/2\pi \sigma^2)e^{-r^2/2\sigma^2}$ is the weight function.

The local stress field is calculated using the Irving-Kirkwood formula as~\cite{IrvingKirkwood}
\begin{equation}
\tensor{\sigma}(\bi r)=-\sum_i m_i \bi v_i \bi v_i\delta(\bi r -\bi r_i)-\sum_{i<j}\bi f_{ij}\bi r_{ij}\int_0^1 ds\, \delta(s\bi r_i+(1-s)\bi r_j-\bi r),
\end{equation}
where the first and second terms denote the kinetic and interaction (configuration) terms, respectively. $\bi f_{ij}$ represents interparticle forces arising from the pair interaction term (the first term in Eq.(1)). In the figures, we also apply coarse-graining to the stress field using the weight function $w(\bi r)$.

\noindent{\bf Identification of half-skyrmions and the bond orientational order parameter}
The Delaunay triangulation shown in Fig.\ref{fig:half-skyrmion} was constructed as follows. First, we must identify the centres of the half-skyrmions. We consider a molecule as a member of a half-skyrmion when $n_z>0.91$. Among such molecules, the nearest neighbour particles (determined by the first peak of the radial distribution function) are defined to be clustered. We define the centre of a mass of clustered molecules as the centre of a half-skyrmion $\bi r_{\rm v}$. The radial distribution function $G_{\rm v}(r)$ and the hexatic correlation function $G_{\rm v6}(r)$ (the angular average is used in this study) are calculated from $\bi r_{\rm v}$, the latter of which is defined as
\begin{equation}
G_{\rm v6}(r)=\av{\psi(0)\psi^*(\bi r)},
\label{eq:G6}
\end{equation}
where the bracket denotes the angular, space, and sample averaging processes, and $\psi(\bi r)=\sum_j\delta(\bi r-\bi r_j)\sum_{k\in {\rm bond}}\exp[6i\theta_{jk}]$ with $\theta_{jk}$ is the bond orientational angle between the displacement vector $\bi r_{{\rm v}k}-\bi r_{{\rm v}j}$ and $x$-axis. $G_{\rm v6}(r)/G_{\rm v}(r)$ is displayed in Fig.\ref{fig:half-skyrmion}a.
In Fig.\ref{fig:half-skyrmion}b, we display the spatial correlation function along with the direction of the bond orientational angle.
We then perform Delaunay triangulation to obtain Fig.\ref{fig:half-skyrmion}f-h.

\vspace{2mm} 
\noindent 
\noindent{\bf Data availability}
Data that support the figures within this paper and the other findings of this study are available from the corresponding authors upon reasonable request.

\vspace{2mm} 
\noindent 
\noindent{\bf Code availability}
The computer codes used in this paper are available from the corresponding authors upon reasonable request.


\begin{thebibliography}{10}
\expandafter\ifx\csname url\endcsname\relax
  \def\url#1{\texttt{#1}}\fi
\expandafter\ifx\csname urlprefix\endcsname\relax\def\urlprefix{URL }\fi
\providecommand{\bibinfo}[2]{#2}
\providecommand{\eprint}[2][]{\url{#2}}

\bibitem{Brazovskii}
\bibinfo{author}{Brazovskii, S.}
\newblock \bibinfo{title}{Phase transition of an isotropic system to a
  nonuniform state}.
\newblock \emph{\bibinfo{journal}{Sov. Phys. JETP}}
  \textbf{\bibinfo{volume}{41}}, \bibinfo{pages}{85--89}
  (\bibinfo{year}{1975}).

\bibitem{Leibler}
\bibinfo{author}{Ruzette, A.-V.} \& \bibinfo{author}{Leibler, L.}
\newblock \bibinfo{title}{Block copolymers in tomorrow's plastics}.
\newblock \emph{\bibinfo{journal}{Nature Mater.}} \textbf{\bibinfo{volume}{4}},
  \bibinfo{pages}{19--31} (\bibinfo{year}{2005}).

\bibitem{NagaosaTokura2013}
\bibinfo{author}{Nagaosa, N.} \& \bibinfo{author}{Tokura, Y.}
\newblock \bibinfo{title}{Topological properties and dynamics of magnetic
  skyrmions}.
\newblock \emph{\bibinfo{journal}{Nature Nanotech.}}
  \textbf{\bibinfo{volume}{8}}, \bibinfo{pages}{899--911}
  (\bibinfo{year}{2013}).

\bibitem{Kawamura}
\bibinfo{author}{Okubo, T.}, \bibinfo{author}{Chung, S.} \&
  \bibinfo{author}{Kawamura, H.}
\newblock \bibinfo{title}{Multiple-$q$ states and the skyrmion lattice of the
  triangular-lattice heisenberg antiferromagnet under magnetic fields}.
\newblock \emph{\bibinfo{journal}{Phys. Rev. Lett.}}
  \textbf{\bibinfo{volume}{108}}, \bibinfo{pages}{017206}
  (\bibinfo{year}{2012}).

\bibitem{Ho-BEC}
\bibinfo{author}{Ho, T.-L.}
\newblock \bibinfo{title}{Spinor {Bose} condensates in optical traps}.
\newblock \emph{\bibinfo{journal}{Phys. Rev. Lett.}}
  \textbf{\bibinfo{volume}{81}}, \bibinfo{pages}{742--745}
  (\bibinfo{year}{1998}).

\bibitem{QHE}
\bibinfo{author}{Sondhi, S.~L.}, \bibinfo{author}{Karlhede, A.},
  \bibinfo{author}{Kivelson, S.~A.} \& \bibinfo{author}{Rezayi, E.~H.}
\newblock \bibinfo{title}{Skyrmions and the crossover from the integer to
  fractional quantum {Hall} effect at small {Zeeman} energies}.
\newblock \emph{\bibinfo{journal}{Phys. Rev. B}} \textbf{\bibinfo{volume}{47}},
  \bibinfo{pages}{16419--16426} (\bibinfo{year}{1993}).

\bibitem{Nahas-polar}
\bibinfo{author}{Nahas, Y.} \emph{et~al.}
\newblock \bibinfo{title}{Discovery of stable skyrmionic state in ferroelectric
  nanocomposites}.
\newblock \emph{\bibinfo{journal}{Nature Commun.}}
  \textbf{\bibinfo{volume}{6}}, \bibinfo{pages}{1--6} (\bibinfo{year}{2015}).

\bibitem{Ramesh-polar2019}
\bibinfo{author}{Das, S.} \emph{et~al.}
\newblock \bibinfo{title}{Observation of room-temperature polar skyrmions}.
\newblock \emph{\bibinfo{journal}{Nature}} \textbf{\bibinfo{volume}{568}},
  \bibinfo{pages}{368--372} (\bibinfo{year}{2019}).

\bibitem{Fukuda2011}
\bibinfo{author}{Fukuda, J.-i.} \& \bibinfo{author}{{\v{Z}}umer, S.}
\newblock \bibinfo{title}{Quasi-two-dimensional skyrmion lattices in a chiral
  nematic liquid crystal}.
\newblock \emph{\bibinfo{journal}{Nature Commun.}}
  \textbf{\bibinfo{volume}{2}}, \bibinfo{pages}{246} (\bibinfo{year}{2011}).

\bibitem{Fukuda2017}
\bibinfo{author}{Nych, A.}, \bibinfo{author}{Fukuda, J.-i.},
  \bibinfo{author}{Ognysta, U.}, \bibinfo{author}{{\v{Z}}umer, S.} \&
  \bibinfo{author}{Mu{\v{s}}evi{\v{c}}, I.}
\newblock \bibinfo{title}{Spontaneous formation and dynamics of half-skyrmions
  in a chiral liquid-crystal film}.
\newblock \emph{\bibinfo{journal}{Nature Phys.}} \textbf{\bibinfo{volume}{13}},
  \bibinfo{pages}{1215--1220} (\bibinfo{year}{2017}).

\bibitem{DePablo}
\bibinfo{author}{Zhang, R.}, \bibinfo{author}{Mozaffari, A.} \&
  \bibinfo{author}{de~Pablo, J.~J.}
\newblock \bibinfo{title}{Autonomous materials systems from active liquid
  crystals}.
\newblock \emph{\bibinfo{journal}{Nature Rev. Mater.}} \bibinfo{pages}{1--17}
  (\bibinfo{year}{2021}).

\bibitem{Shankar-active}
\bibinfo{author}{Shankar, S.}, \bibinfo{author}{Souslov, A.},
  \bibinfo{author}{Bowick, M.~J.}, \bibinfo{author}{Marchetti, M.~C.} \&
  \bibinfo{author}{Vitelli, V.}
\newblock \bibinfo{title}{Topological active matter}  (\bibinfo{year}{2020}).
\newblock arXiv:\eprint{2010.00364}.

\bibitem{Tokura2021review}
\bibinfo{author}{Tokura, Y.} \& \bibinfo{author}{Kanazawa, N.}
\newblock \bibinfo{title}{Magnetic skyrmion materials}.
\newblock \emph{\bibinfo{journal}{Chem. Rev.}} \textbf{\bibinfo{volume}{121}},
  \bibinfo{pages}{2857--2897} (\bibinfo{year}{2021}).

\bibitem{Smalyukh}
\bibinfo{author}{Foster, D.} \emph{et~al.}
\newblock \bibinfo{title}{Two-dimensional skyrmion bags in liquid crystals and
  ferromagnets}.
\newblock \emph{\bibinfo{journal}{Nature Phys.}} \textbf{\bibinfo{volume}{15}},
  \bibinfo{pages}{655--659} (\bibinfo{year}{2019}).

\bibitem{pinning}
\bibinfo{author}{Reichhardt, C.}, \bibinfo{author}{Reichhardt, C. J.~O.} \&
  \bibinfo{author}{Milosevic, M.~V.}
\newblock \bibinfo{title}{Statics and dynamics of skyrmions interacting with
  pinning: A review}  (\bibinfo{year}{2021}).
\newblock arXiv:\eprint{2102.10464}.

\bibitem{deGennesProst}
\bibinfo{author}{De~Gennes, P.-G.} \& \bibinfo{author}{Prost, J.}
\newblock \emph{\bibinfo{title}{The Physics of Liquid Crystals}}
  (\bibinfo{publisher}{Oxford university press}, \bibinfo{year}{1993}).

\bibitem{Mermin}
\bibinfo{author}{Wright, D.~C.} \& \bibinfo{author}{Mermin, N.~D.}
\newblock \bibinfo{title}{Crystalline liquids: the blue phases}.
\newblock \emph{\bibinfo{journal}{Rev. Mod. Phys.}}
  \textbf{\bibinfo{volume}{61}}, \bibinfo{pages}{385--432}
  (\bibinfo{year}{1989}).

\bibitem{Tschierske2014}
\bibinfo{author}{Dressel, C.}, \bibinfo{author}{Reppe, T.},
  \bibinfo{author}{Prehm, M.}, \bibinfo{author}{Brautzsch, M.} \&
  \bibinfo{author}{Tschierske, C.}
\newblock \bibinfo{title}{Chiral self-sorting and amplification in isotropic
  liquids of achiral molecules}.
\newblock \emph{\bibinfo{journal}{Nature Chem.}} \textbf{\bibinfo{volume}{6}},
  \bibinfo{pages}{971--977} (\bibinfo{year}{2014}).

\bibitem{Yeomans}
\bibinfo{author}{Metselaar, L.}, \bibinfo{author}{Doostmohammadi, A.} \&
  \bibinfo{author}{Yeomans, J.~M.}
\newblock \bibinfo{title}{Topological states in chiral active matter: Dynamic
  blue phases and active half-skyrmions}.
\newblock \emph{\bibinfo{journal}{J. Chem. Phys.}}
  \textbf{\bibinfo{volume}{150}}, \bibinfo{pages}{064909}
  (\bibinfo{year}{2019}).

\bibitem{Roessler}
\bibinfo{author}{R{\"{o}}{\ss}ler, U.~K.}, \bibinfo{author}{Bogdanov, A.~N.} \&
  \bibinfo{author}{Pfleiderer, C.}
\newblock \bibinfo{title}{Spontaneous skyrmion ground states in magnetic
  metals}.
\newblock \emph{\bibinfo{journal}{Nature}} \textbf{\bibinfo{volume}{442}},
  \bibinfo{pages}{797--801} (\bibinfo{year}{2006}).

\bibitem{Skyrmion-MOF}
\bibinfo{author}{Wolpert, E.}, \bibinfo{author}{Coudert, F.-X.} \&
  \bibinfo{author}{Goodwin, A.}
\newblock \bibinfo{title}{Skyrmion lattices in chiral metal-organic frameworks}
   (\bibinfo{year}{2020}).
\newblock \bibinfo{note}{ChemRxiv.12515594}.

\bibitem{Viedma2006}
\bibinfo{author}{Klussmann, M.} \emph{et~al.}
\newblock \bibinfo{title}{Thermodynamic control of asymmetric amplification in
  amino acid catalysis}.
\newblock \emph{\bibinfo{journal}{Nature}} \textbf{\bibinfo{volume}{441}},
  \bibinfo{pages}{621--623} (\bibinfo{year}{2006}).

\bibitem{softcrystal}
\bibinfo{author}{Kato, M.}, \bibinfo{author}{Ito, H.},
  \bibinfo{author}{Hasegawa, M.} \& \bibinfo{author}{Ishii, K.}
\newblock \bibinfo{title}{Soft crystals: Flexible response systems with high
  structural order}.
\newblock \emph{\bibinfo{journal}{Chem. Eur. J.}}
  \textbf{\bibinfo{volume}{25}}, \bibinfo{pages}{5105--5112}
  (\bibinfo{year}{2019}).

\bibitem{Dogic2017}
\bibinfo{author}{Siavashpouri, M.} \emph{et~al.}
\newblock \bibinfo{title}{Molecular engineering of chiral colloidal liquid
  crystals using {DNA} origami}.
\newblock \emph{\bibinfo{journal}{Nature Mater.}}
  \textbf{\bibinfo{volume}{16}}, \bibinfo{pages}{849--856}
  (\bibinfo{year}{2017}).

\bibitem{Dijkstra2016}
\bibinfo{author}{Dussi, S.} \& \bibinfo{author}{Dijkstra, M.}
\newblock \bibinfo{title}{Entropy-driven formation of chiral nematic phases by
  computer simulations}.
\newblock \emph{\bibinfo{journal}{Nature Commun.}}
  \textbf{\bibinfo{volume}{7}}, \bibinfo{pages}{11175} (\bibinfo{year}{2016}).

\bibitem{chemistry}
\bibinfo{author}{Ruiz-Mirazo, K.}, \bibinfo{author}{Briones, C.} \&
  \bibinfo{author}{de~la Escosura, A.}
\newblock \bibinfo{title}{Prebiotic systems chemistry: New perspectives for the
  origins of life}.
\newblock \emph{\bibinfo{journal}{Chem. Rev.}} \textbf{\bibinfo{volume}{114}},
  \bibinfo{pages}{285--366} (\bibinfo{year}{2014}).

\bibitem{emergent}
\bibinfo{author}{Schulz, T.} \emph{et~al.}
\newblock \bibinfo{title}{Emergent electrodynamics of skyrmions in a chiral
  magnet}.
\newblock \emph{\bibinfo{journal}{Nature Phys.}} \textbf{\bibinfo{volume}{8}},
  \bibinfo{pages}{301--304} (\bibinfo{year}{2012}).

\bibitem{Nagaosa-emergent}
\bibinfo{author}{Nagaosa, N.}
\newblock \bibinfo{title}{Emergent electromagnetism in condensed matter}.
\newblock \emph{\bibinfo{journal}{Proc. Jpn. Acad., Ser. B}}
  \textbf{\bibinfo{volume}{95}}, \bibinfo{pages}{278--289}
  (\bibinfo{year}{2019}).

\bibitem{Eshelby}
\bibinfo{author}{Eshelby, J.~D.}
\newblock \bibinfo{title}{The determination of the elastic field of an
  ellipsoidal inclusion, and related problems}.
\newblock \emph{\bibinfo{journal}{Proc. R. Soc. Lond. A}}
  \textbf{\bibinfo{volume}{241}}, \bibinfo{pages}{376--396}
  (\bibinfo{year}{1957}).

\bibitem{Lemaitre}
\bibinfo{author}{Maloney, C.~E.} \& \bibinfo{author}{Lema\^{\i}tre, A.}
\newblock \bibinfo{title}{Amorphous systems in athermal, quasistatic shear}.
\newblock \emph{\bibinfo{journal}{Phys. Rev. E}} \textbf{\bibinfo{volume}{74}},
  \bibinfo{pages}{016118} (\bibinfo{year}{2006}).

\bibitem{Terentjev}
\bibinfo{author}{Warner, M.} \& \bibinfo{author}{Terentjev, E.~M.}
\newblock \emph{\bibinfo{title}{Liquid Crystal Elastomers}}
  (\bibinfo{publisher}{Oxford University Press}, \bibinfo{year}{2003}).

\bibitem{KT1973}
\bibinfo{author}{Kosterlitz, J.~M.} \& \bibinfo{author}{Thouless, D.~J.}
\newblock \bibinfo{title}{Ordering, metastability and phase transitions in
  two-dimensional systems}.
\newblock \emph{\bibinfo{journal}{J. Phys. C: Solid State Phys.}}
  \textbf{\bibinfo{volume}{6}}, \bibinfo{pages}{1181--1203}
  (\bibinfo{year}{1973}).

\bibitem{NelsonHalperin}
\bibinfo{author}{Nelson, D.~R.} \& \bibinfo{author}{Halperin, B.~I.}
\newblock \bibinfo{title}{Dislocation-mediated melting in two dimensions}.
\newblock \emph{\bibinfo{journal}{Phys. Rev. B}} \textbf{\bibinfo{volume}{19}},
  \bibinfo{pages}{2457--2484} (\bibinfo{year}{1979}).

\bibitem{Young}
\bibinfo{author}{Young, A.~P.}
\newblock \bibinfo{title}{Melting and the vector coulomb gas in two
  dimensions}.
\newblock \emph{\bibinfo{journal}{Phys. Rev. B}} \textbf{\bibinfo{volume}{19}},
  \bibinfo{pages}{1855--1866} (\bibinfo{year}{1979}).

\bibitem{Krauth2015}
\bibinfo{author}{Kapfer, S.~C.} \& \bibinfo{author}{Krauth, W.}
\newblock \bibinfo{title}{Two-dimensional melting: From liquid-hexatic
  coexistence to continuous transitions}.
\newblock \emph{\bibinfo{journal}{Phys. Rev. Lett.}}
  \textbf{\bibinfo{volume}{114}}, \bibinfo{pages}{035702}
  (\bibinfo{year}{2015}).

\bibitem{Huang}
\bibinfo{author}{Huang, P.} \emph{et~al.}
\newblock \bibinfo{title}{Melting of a skyrmion lattice to a skyrmion liquid
  via a hexatic phase}.
\newblock \emph{\bibinfo{journal}{Nature Nanotech.}}
  \textbf{\bibinfo{volume}{15}}, \bibinfo{pages}{761--767}
  (\bibinfo{year}{2020}).

\bibitem{Zazvorka}
\bibinfo{author}{Z{\'{a}}zvorka, J.} \emph{et~al.}
\newblock \bibinfo{title}{Skyrmion lattice phases in thin film multilayer}.
\newblock \emph{\bibinfo{journal}{Adv. Funct. Mater.}}
  \textbf{\bibinfo{volume}{30}}, \bibinfo{pages}{2004037}
  (\bibinfo{year}{2020}).

\bibitem{PMA}
\bibinfo{author}{Dieny, B.} \& \bibinfo{author}{Chshiev, M.}
\newblock \bibinfo{title}{Perpendicular magnetic anisotropy at transition
  metal/oxide interfaces and applications}.
\newblock \emph{\bibinfo{journal}{Rev. Mod. Phys.}}
  \textbf{\bibinfo{volume}{89}}, \bibinfo{pages}{025008}
  (\bibinfo{year}{2017}).

\bibitem{Dogic}
\bibinfo{author}{Sharma, P.}, \bibinfo{author}{Ward, A.},
  \bibinfo{author}{Gibaud, T.}, \bibinfo{author}{Hagan, M.~F.} \&
  \bibinfo{author}{Dogic, Z.}
\newblock \bibinfo{title}{Hierarchical organization of chiral rafts in
  colloidal membranes}.
\newblock \emph{\bibinfo{journal}{Nature}} \textbf{\bibinfo{volume}{513}},
  \bibinfo{pages}{77--80} (\bibinfo{year}{2014}).

\bibitem{MacLachlan}
\bibinfo{author}{Tran, A.}, \bibinfo{author}{Boott, C.~E.} \&
  \bibinfo{author}{MacLachlan, M.~J.}
\newblock \bibinfo{title}{Understanding the self-assembly of cellulose
  nanocrystals--toward chiral photonic materials}.
\newblock \emph{\bibinfo{journal}{Adv. Mater.}} \textbf{\bibinfo{volume}{32}},
  \bibinfo{pages}{1905876} (\bibinfo{year}{2020}).

\bibitem{Zerrouki}
\bibinfo{author}{Zerrouki, D.}, \bibinfo{author}{Baudry, J.},
  \bibinfo{author}{Pine, D.}, \bibinfo{author}{Chaikin, P.} \&
  \bibinfo{author}{Bibette, J.}
\newblock \bibinfo{title}{Chiral colloidal clusters}.
\newblock \emph{\bibinfo{journal}{Nature}} \textbf{\bibinfo{volume}{455}},
  \bibinfo{pages}{380--382} (\bibinfo{year}{2008}).

\bibitem{Singh}
\bibinfo{author}{Singh, G.} \emph{et~al.}
\newblock \bibinfo{title}{Self-assembly of magnetite nanocubes into helical
  superstructures}.
\newblock \emph{\bibinfo{journal}{Science}} \textbf{\bibinfo{volume}{345}},
  \bibinfo{pages}{1149--1153} (\bibinfo{year}{2014}).

\bibitem{Nanorod}
\bibinfo{author}{Lan, X.} \emph{et~al.}
\newblock \bibinfo{title}{Au nanorod helical superstructures with designed
  chirality}.
\newblock \emph{\bibinfo{journal}{J. Am. Chem. Soc.}}
  \textbf{\bibinfo{volume}{137}}, \bibinfo{pages}{457--462}
  (\bibinfo{year}{2015}).

\bibitem{Rosi}
\bibinfo{author}{Mokashi-Punekar, S.}, \bibinfo{author}{Zhou, Y.},
  \bibinfo{author}{Brooks, S.~C.} \& \bibinfo{author}{Rosi, N.~L.}
\newblock \bibinfo{title}{Construction of chiral, helical nanoparticle
  superstructures: Progress and prospects}.
\newblock \emph{\bibinfo{journal}{Adv. Mater.}} \textbf{\bibinfo{volume}{32}},
  \bibinfo{pages}{1905975} (\bibinfo{year}{2020}).

\bibitem{Sacanna2021}
\bibinfo{author}{Hueckel, T.}, \bibinfo{author}{Hocky, G.~M.} \&
  \bibinfo{author}{Sacanna, S.}
\newblock \bibinfo{title}{Total synthesis of colloidal matter}.
\newblock \emph{\bibinfo{journal}{Nature Rev. Mater.}} \bibinfo{pages}{1--17}
  (\bibinfo{year}{2021}).

\bibitem{AFM}
\bibinfo{author}{Garcia, R.}
\newblock \bibinfo{title}{Nanomechanical mapping of soft materials with the
  atomic force microscope: methods{,} theory and applications}.
\newblock \emph{\bibinfo{journal}{Chem. Soc. Rev.}}
  \textbf{\bibinfo{volume}{49}}, \bibinfo{pages}{5850--5884}
  (\bibinfo{year}{2020}).

\bibitem{Spaepen}
\bibinfo{author}{Schall, P.}, \bibinfo{author}{Cohen, I.},
  \bibinfo{author}{Weitz, D.~A.} \& \bibinfo{author}{Spaepen, F.}
\newblock \bibinfo{title}{Visualizing dislocation nucleation by indenting
  colloidal crystals}.
\newblock \emph{\bibinfo{journal}{Nature}} \textbf{\bibinfo{volume}{440}},
  \bibinfo{pages}{319--323} (\bibinfo{year}{2006}).

\bibitem{Takae2018antiferro}
\bibinfo{author}{Takae, K.} \& \bibinfo{author}{Tanaka, H.}
\newblock \bibinfo{title}{Self-organization into ferroelectric and
  antiferroelectric crystals via the interplay between particle shape and
  dipolar interaction}.
\newblock \emph{\bibinfo{journal}{Proc. Natl. Acad. Sci.}}
  \textbf{\bibinfo{volume}{115}}, \bibinfo{pages}{9917--9922}
  (\bibinfo{year}{2018}).

\bibitem{Tokura2015strain}
\bibinfo{author}{Shibata, K.} \emph{et~al.}
\newblock \bibinfo{title}{Large anisotropic deformation of skyrmions in
  strained crystal}.
\newblock \emph{\bibinfo{journal}{Nature nanotechnology}}
  \textbf{\bibinfo{volume}{10}}, \bibinfo{pages}{589--592}
  (\bibinfo{year}{2015}).

\bibitem{Wang2019-mechanical}
\bibinfo{author}{Wang, J.}
\newblock \bibinfo{title}{Mechanical control of magnetic order: From phase
  transition to skyrmions}.
\newblock \emph{\bibinfo{journal}{Annu. Rev. Mater. Res.}}
  \textbf{\bibinfo{volume}{49}}, \bibinfo{pages}{361--388}
  (\bibinfo{year}{2019}).

\bibitem{Takae2014origlass}
\bibinfo{author}{Takae, K.} \& \bibinfo{author}{Onuki, A.}
\newblock \bibinfo{title}{Orientational glass in mixtures of elliptic and
  circular particles: Structural heterogeneities, rotational dynamics, and
  rheology}.
\newblock \emph{\bibinfo{journal}{Phys. Rev. E}} \textbf{\bibinfo{volume}{89}},
  \bibinfo{pages}{022308} (\bibinfo{year}{2014}).

\bibitem{Takae2017relaxor}
\bibinfo{author}{Takae, K.} \& \bibinfo{author}{Onuki, A.}
\newblock \bibinfo{title}{Ferroelectric glass of spheroidal dipoles with
  impurities: Polar nanoregions, response to applied electric field, and
  ergodicity breakdown}.
\newblock \emph{\bibinfo{journal}{J. Phys.: Condens. Matter}}
  \textbf{\bibinfo{volume}{29}}, \bibinfo{pages}{165401}
  (\bibinfo{year}{2017}).

\bibitem{Meer}
\bibinfo{author}{van~der Meer, B.~W.}, \bibinfo{author}{Vertogen, G.},
  \bibinfo{author}{Dekker, A.~J.} \& \bibinfo{author}{Ypma, J. G.~J.}
\newblock \bibinfo{title}{A molecular-statistical theory of the
  temperature-dependent pitch in cholesteric liquid crystals}.
\newblock \emph{\bibinfo{journal}{J. Chem. Phys.}}
  \textbf{\bibinfo{volume}{65}}, \bibinfo{pages}{3935--3943}
  (\bibinfo{year}{1976}).

\bibitem{Memmer}
\bibinfo{author}{Memmer, R.}, \bibinfo{author}{Kuball, H.-G.} \&
  \bibinfo{author}{Schönhofer, A.}
\newblock \bibinfo{title}{Computer simulation of chiral liquid crystal phases.
  i. the polymorphism of the chiral gay-berne fluid}.
\newblock \emph{\bibinfo{journal}{Liq. Cryst.}} \textbf{\bibinfo{volume}{15}},
  \bibinfo{pages}{345--360} (\bibinfo{year}{1993}).

\bibitem{Buka}
\bibinfo{author}{Buka, A.} \& \bibinfo{author}{De~Jeu, W.}
\newblock \bibinfo{title}{Diamagnetism and orientational order of nematic
  liquid crystals}.
\newblock \emph{\bibinfo{journal}{J. Phys. (Paris)}}
  \textbf{\bibinfo{volume}{43}}, \bibinfo{pages}{361--367}
  (\bibinfo{year}{1982}).

\bibitem{Allen}
\bibinfo{author}{Allen, M.~P.} \& \bibinfo{author}{Tildesley, D.~J.}
\newblock \emph{\bibinfo{title}{{Computer Simulation of Liquids}}}
  (\bibinfo{publisher}{Oxford university press}, \bibinfo{year}{1989}).

\bibitem{IrvingKirkwood}
\bibinfo{author}{Irving, J.~H.} \& \bibinfo{author}{Kirkwood, J.~G.}
\newblock \bibinfo{title}{The statistical mechanical theory of transport
  processes. iv. the equations of hydrodynamics}.
\newblock \emph{\bibinfo{journal}{J. Chem. Phys.}}
  \textbf{\bibinfo{volume}{18}}, \bibinfo{pages}{817--829}
  (\bibinfo{year}{1950}).

\end{thebibliography}

\vspace{5mm}
\noindent 
{\bf Acknowledgments} K.T. was supported by JSPS KAKENHI Grant Numbers JP17H06375 and JP20H05619. T.K. was supported by JSPS KAKENHI Grant Numbers JP20H00128, JP20H05157, JP19K03767, and JP18H01188.

\vspace{2mm} 
\noindent 
{\bf Author contributions.} K.T. and T.K. conceived the project, K.T. performed numerical simulations, K.T. and T.K. discussed the results, and K.T. wrote the manuscript.

\vspace{2mm} 
\noindent 
{\bf Competing interests.} 
The authors declare no competing financial interests.

\vspace{2mm} 
\noindent 
{\bf Corresponding authors}
Correspondence and requests for materials should be addressed to K.T. or T.K.

\clearpage

\renewcommand{\figurename}{{\bf Extended Data Fig.~}}
\setcounter{figure}{0}

\begin{figure}[t!]
\centering
\includegraphics[width=15cm]{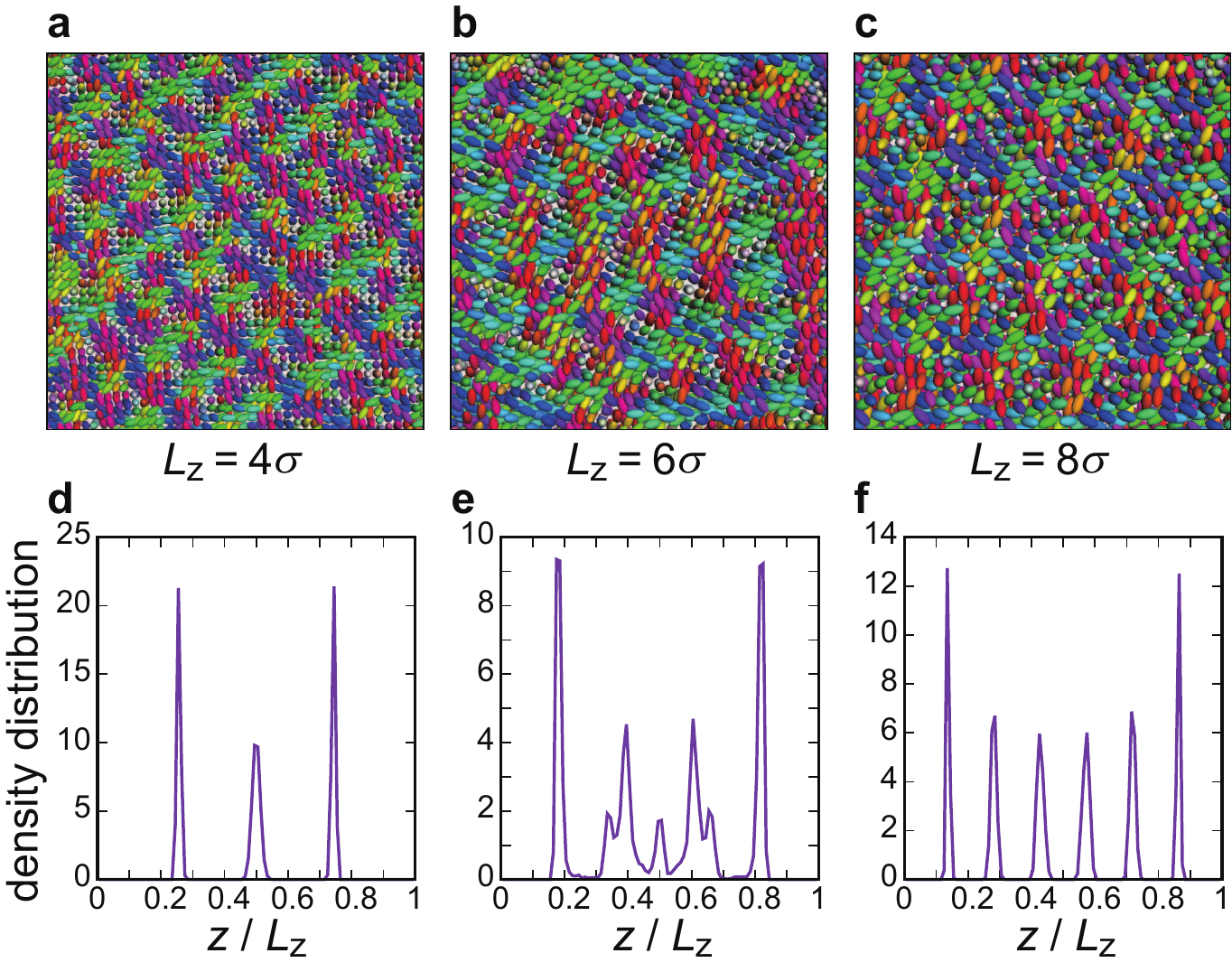}
\caption{
{\bf Quasi-two-dimensional simulations with controllable layer thicknesses for $K_2=50$ and $q_0=0.3$.}
{\bf a-c.} Top view of the particle configuration for ({\bf a}) $L_z=4\sigma$, ({\bf b}) $L_z=6\sigma$, and ({\bf c}) $L_z=8\sigma$. A uniform half-skyrmion structure along $z$-direction emerges in ({\bf a}) and ({\bf b}). In ({\bf c}), many half-skyrmion structures do not penetrate into the system and in-plane zig-zag structures develop.
{\bf d-f.} Density profiles showing that ({\bf d}) three distinct layers are formed for $L_z=4\sigma$, whereas ({\bf e}) four- and five-layer regions are heterogeneously distributed for $L_z=6\sigma$. Particles align parallel (perpendicular) to the $z$-direction for the former (latter) profiles and produce large steric repulsions along the particle orientation. Therefore, the skyrmion cores have a smaller number of layers. ({\bf f}) $L_z=8\sigma$; here, most of the inner particles are aligned perpendicularly and the density profile again exhibits perfect layering.
}
\label{fig:Q2d}
\end{figure}

\begin{figure}[t!]
\centering
\includegraphics[width=15cm]{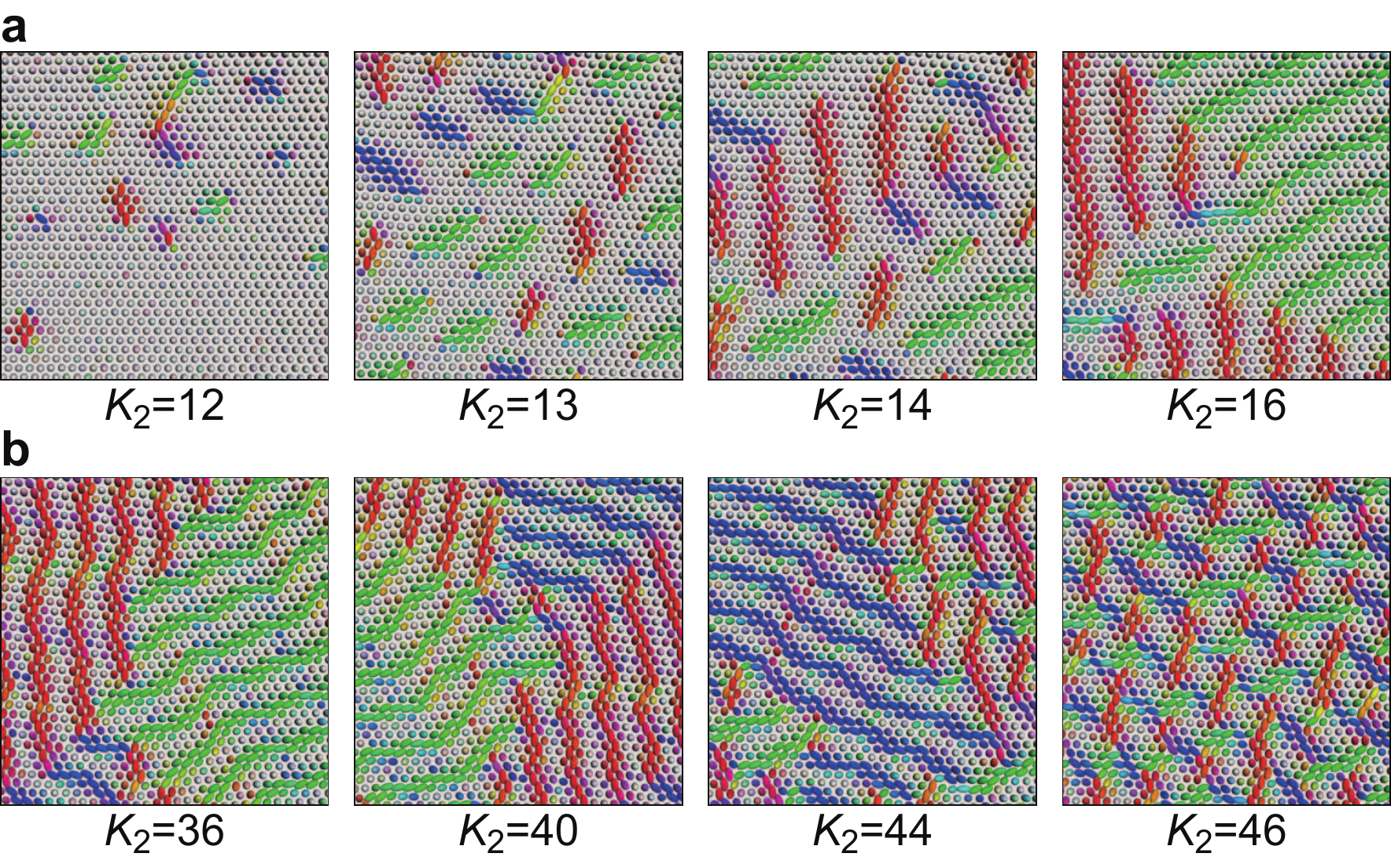}
\caption{
{\bf Low-temperature phase behaviour at $q_0=0.3$.}
{\bf a.} As the twist rigidity $K_2$ increases, short chains that are perpendicularly aligned grow ($K_2=12$). At $K_2=13$, these chains exhibit positional ordering. They are connected to form a lamellar structure at larger rigidities ($K_2=14$ and 16) to realise a helical phase.
{\bf b.} For the twist rigidity close to the phase boundary between the helical and half-skyrmion phases, the lamellar structures undulate as $K_2$ increases. This undulation eventually leads to the formation of skyrmions, while skyrmions nucleate at the domain interface.
}
\label{fig:q03}
\end{figure}

\begin{figure}[t!]
\centering
\includegraphics[width=15cm]{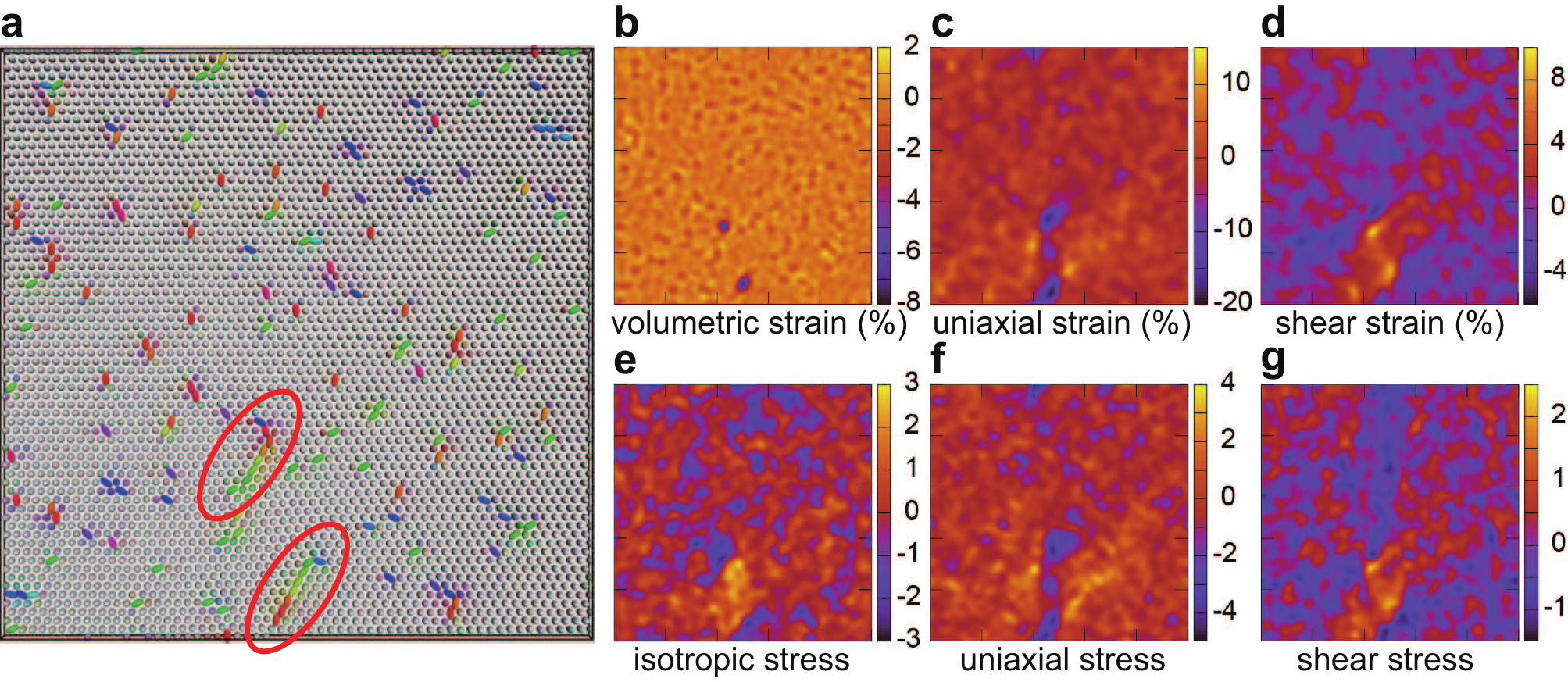}
\caption{
{\bf Emergent elastic field of a uniform state.}
{\bf a.} Particle configuration of the uniform state corresponding to Fig.\ref{fig:phase}e, where $K_2=30$ and $q_0=0.15$. Because there is a small but finite twist interaction, orientational defects develop. Two long orientational defects are indicated by the open red ellipses.
{\bf b-d.} Strain field of this configuration, in which a large strain arises around the long defects for all components.
{\bf e-g.} Stress field of this configuration, showing a similar spatial pattern to the corresponding strain field.
Both the strain and stress fields exhibit a long-range spatial correlation inside of the two long defects.
}
\label{fig:Em-uni}
\end{figure}

\begin{figure}[t!]
\centering
\includegraphics[width=15cm]{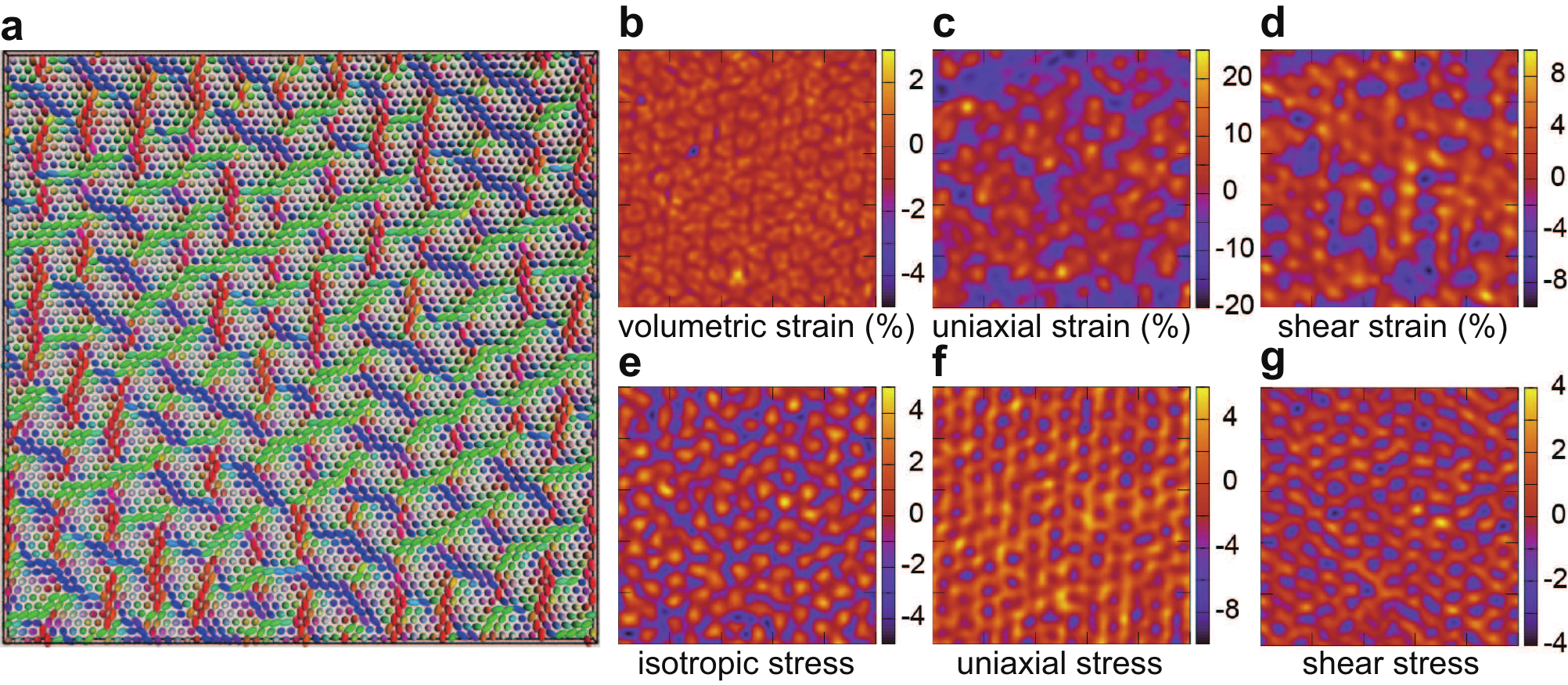}
\caption{
{\bf Emergent elastic field of a half-skyrmion state.}
{\bf a.} Particle configuration of the half-skyrmion state corresponding to Fig.\ref{fig:phase}g where $K_2=60$ and $q_0=0.25$.
{\bf b-d.} Strain field of this configuration. A large spontaneous strain is induced by the formation of the half-skyrmion phase and is distributed heterogeneously.
{\bf e-g.} Stress field of this configuration, showing a characteristic hexagonal pattern due to the formation of half-skyrmions.
}
\label{fig:Em-sky}
\end{figure}

\begin{figure}[t!]
\centering
\includegraphics[width=7cm]{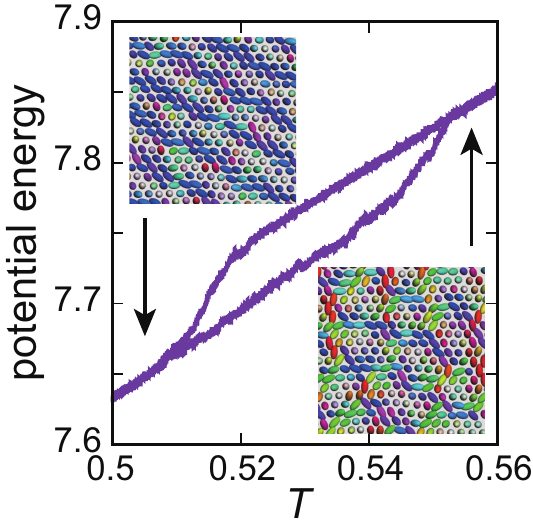}
\caption{
{\bf Thermal phase transition between the helical and half-skyrmion phases.}
The temperature dependence of the potential energy is displayed, wherein a thermal hysteresis is observed as the temperature is slowly varied via $dT/dt=10^{-6}$. The inset shows the real spatial structures at each phase.
It is confirmed that both the helical and half-skyrmion states are stable over the entire computational duration when the temperature is fixed $0.525\le T\le 0.540$. At $T=0.52$ and $0.545$, the metastable half-skyrmion and helical states gradually transform with a characteristic relaxation time of $\sim 10^4$. We confirm that the half-skyrmion phase is a half-skyrmion liquid state without a long-range translational and hexatic order in the position of the half-skyrmions.
$K_2=40$ and $q_0=0.3$ in this figure.
}
\label{fig:helical}
\end{figure}

\begin{figure}[t!]
\centering
\includegraphics[width=15cm]{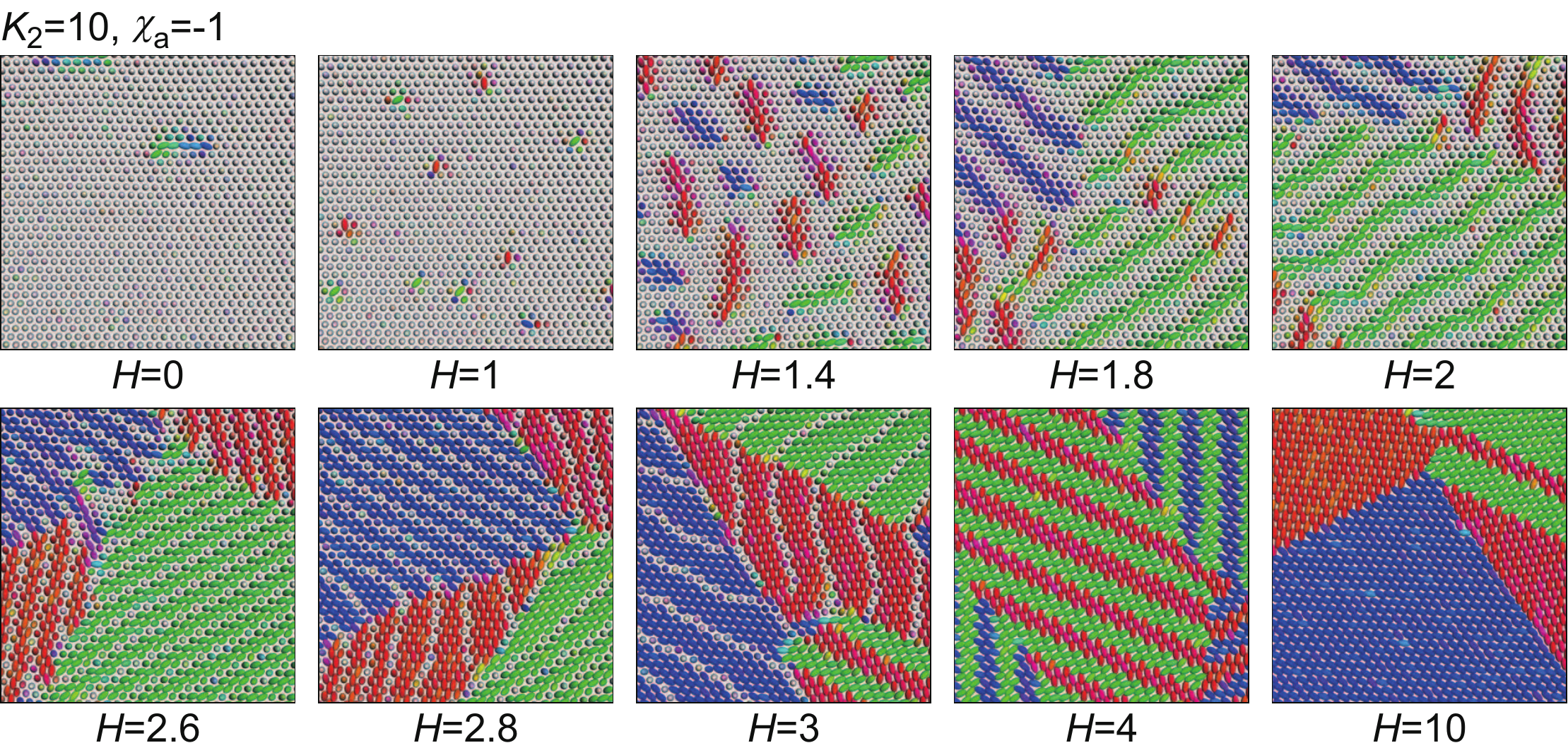}
\caption{
{\bf Magnetic field effects for a uniform state ($K_2=10$) with negative magnetic anisotropy under the field-cooling condition.}
The particle alignment becomes perpendicular for larger fields such that a helical state is formed in the upper row. Until $H=2$,  
the phase behaviour resembles that achieved by increasing $K_2$ in Extended Data Figure \ref{fig:q03}. For larger fields, however, the helical pattern transforms into a striped pattern, and zig-zag and uniform two-dimensional structures are eventually formed for $H=4$ and $H=10$, respectively.
}
\label{fig:10}
\end{figure}

\begin{figure}[t!]
\centering
\includegraphics[width=15cm]{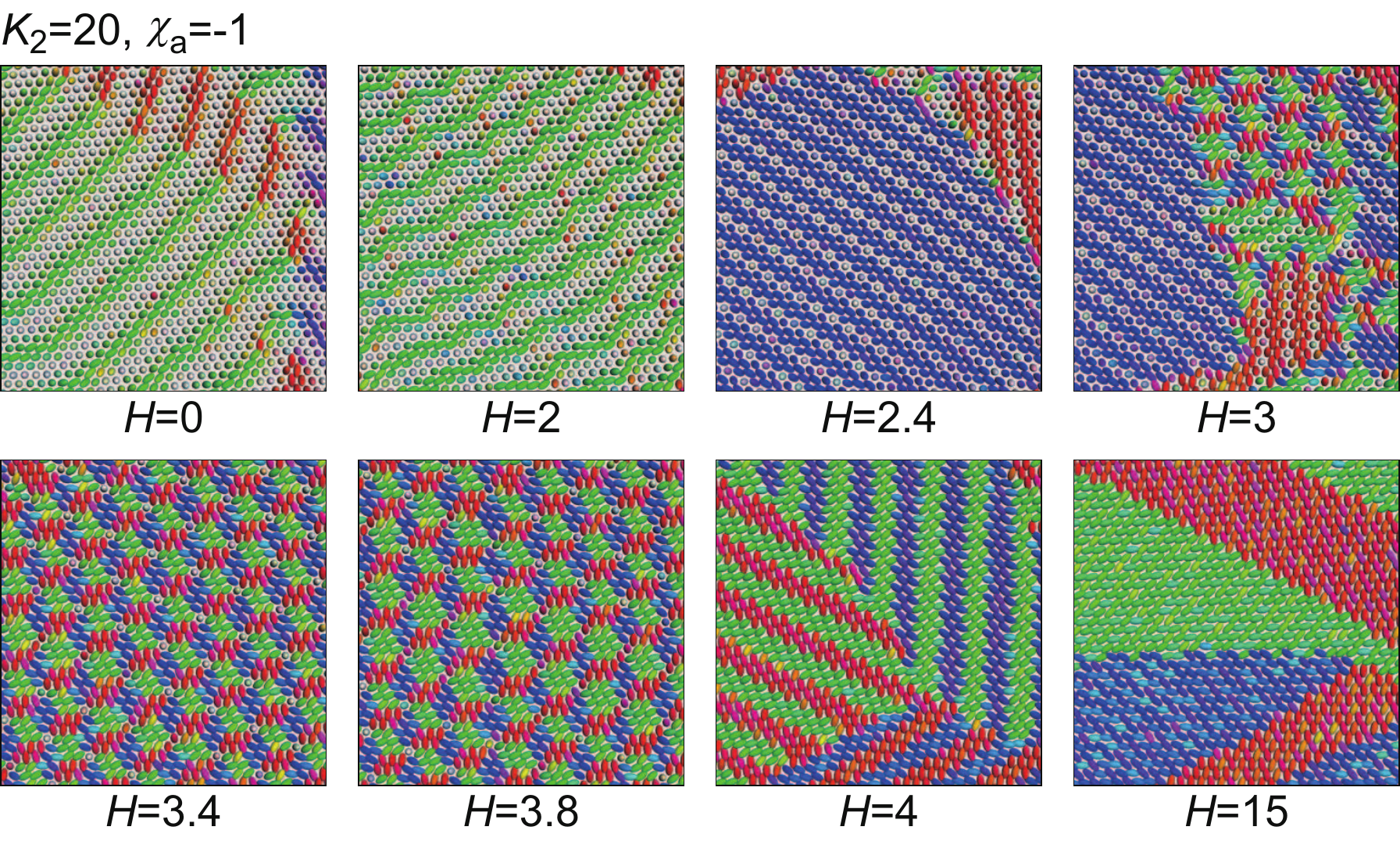}
\caption{
{\bf Magnetic field effects for a helical state ($K_2=20$) with negative magnetic anisotropy under the field-cooling condition.}
The helical state transforms into a striped state, as in the previous case. A half-skyrmion phase emerges under strong fields. This tendency resembles magnetic skyrmion systems, although the sign of the magnetic anisotropy differs. The half-skyrmion states eventually transform into zig-zag and uniform two-dimensional structures.
}
\label{fig:20}
\end{figure}

\begin{figure}[t!]
\centering
\includegraphics[width=15cm]{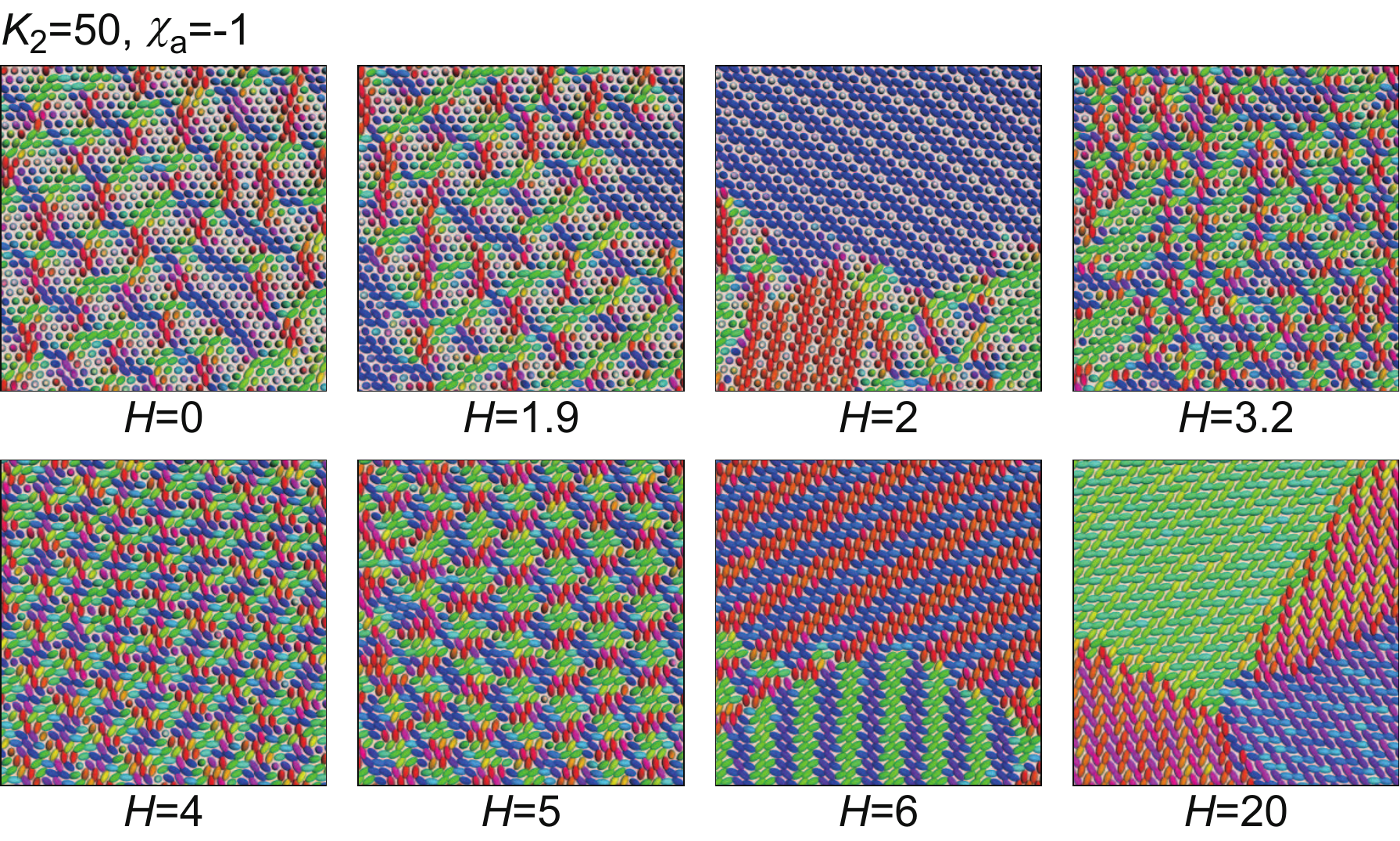}
\caption{
{\bf Details of the magnetic field effects for a half-skyrmion state ($K_2=50$) with a negative magnetic anisotropy corresponding to Fig.\ref{fig:field}.}
Once the striped state has been realised ($H=2$), almost the same phase behaviour appears as in the helical case (Extended Data Figure \ref{fig:20}).
}
\label{fig:50}
\end{figure}

\end{document}